\documentclass{aa}  
\usepackage{natbib}
\bibpunct{(}{)}{;}{a}{}{,} 
\usepackage{graphicx}
\usepackage{txfonts}
\usepackage{hyperref}
\newcommand{\orcit}[1]{\protect\href{https://orcid.org/#1}{\protect\includegraphics[width=8pt]{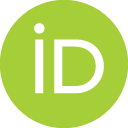}}}

\def\gaia{\textit{Gaia}\xspace}
\def\wise{\textit{WISE}\xspace}
\def\gmag{$G$\xspace}
\def\gbp{\mbox{$G_{\rm BP}$}\xspace}
\def\grp{\mbox{$G_{\rm RP}$}\xspace}

\def\gmr{\mbox{$G-G_{\rm RP}$}\xspace}
\def\bmg{\mbox{$G_{\rm BP}-G$}\xspace}

\usepackage{color}
\usepackage[normalem]{ulem} 

\begin{document}

   \title{\gaia Data Release 3:}

   \subtitle{The first \gaia catalogue of variable AGN}

\author{Maria I.~Carnerero\orcit{0000-0001-5843-5515}\inst{\ref{inst-it-to}}
\and
Claudia M.~Raiteri\orcit{0000-0003-1784-2784}\inst{\ref{inst-it-to}}
\and 
Lorenzo~Rimoldini\orcit{0000-0002-0306-585X}\inst{\ref{inst-ch-eco}}
\and
Deborah~Busonero\orcit{0000-0002-3903-7076}\inst{\ref{inst-it-to}}
\and
Enrico~Licata\orcit{0000-0002-5203-0135}\inst{\ref{inst-it-to}}
\and \\
Nami~Mowlavi\orcit{0000-0003-1578-6993}\inst{\ref{inst-ch-eco},\ref{inst-ch-obs}}
\and
Isabelle~Lecoeur-Ta\"ibi\orcit{0000-0003-0029-8575}\inst{\ref{inst-ch-eco}}
\and
Marc~Audard\orcit{0000-0003-4721-034X}\inst{\ref{inst-ch-eco},\ref{inst-ch-obs}}
\and
Berry~Holl\orcit{0000-0001-6220-3266}\inst{\ref{inst-ch-eco},\ref{inst-ch-obs}}
\and
Panagiotis~Gavras\orcit{0000-0002-4383-4836}\inst{\ref{inst-es-rhea}}
\and
Krzysztof~Nienartowicz\orcit{0000-0001-5415-0547}\inst{\ref{inst-ch-sed}}
\and
Gr\'{e}gory~Jevardat~de~Fombelle\inst{\ref{inst-ch-eco}}
\and
Ruth~Carballo\orcit{0000-0001-7412-2498}\inst{\ref{inst-es}}
\and
Gisella~Clementini\orcit{0000-0001-9206-9723}\inst{\ref{inst-it-bo}}
\and 
Ludovic Delchambre\orcit{0000-0003-2559-408X}\inst{\ref{inst-be}}
\and
Sergei~Klioner\orcit{0000-0003-4682-7831}\inst{\ref{inst-de-obs}}
\and
Mario~G.~Lattanzi\orcit{0000-0003-0429-7748}\inst{\ref{inst-it-to}}
\and
Laurent~Eyer\orcit{0000-0002-0182-8040}\inst{\ref{inst-ch-obs}}
}

\institute{
INAF - Osservatorio Astrofisico di Torino, Via Osservatorio 20, I-10025 Pino Torinese, Italy\label{inst-it-to} \\
\email{maria.carnerero@inaf.it, claudia.raiteri@inaf.it}
\and
Department of Astronomy, University of Geneva, Chemin d'Ecogia 16, CH-1290 Versoix, Switzerland\label{inst-ch-eco}
\and
Department of Astronomy, University of Geneva, Chemin Pegasi 51, CH-1290 Versoix, Switzerland\label{inst-ch-obs}
\and
RHEA for European Space Agency (ESA), Camino bajo del Castillo, s/n, Urbanizacion Villafranca del Castillo, Villanueva de la Cañada, E-28692 Madrid, Spain\label{inst-es-rhea}
\and 
Sednai Sàrl, Geneva, Switzerland\label{inst-ch-sed}
\and
 Dpto. de Matem\'{a}tica Aplicada y Ciencias de la Computaci\'{o}n, Univ. de Cantabria, ETS Ingenieros de Caminos, Canales y Puertos, Avda. de los Castros s/n, 39005 Santander, Spain\label{inst-es}
\and 
INAF - Osservatorio di Astrofisica e Scienza dello Spazio di Bologna, Via Gobetti 93/3, I-40129 Bologna, Italy\label{inst-it-bo}
\and
Lohrmann Observatory, Technische Universit\"at Dresden, Mommsenstra{\ss}e 13, D-01062 Dresden, Germany\label{inst-de-obs}
\and
Space sciences, Technologies \& Astrophysics Research (STAR) Institute,
Institute of Astrophysics and Geophysics,
University of Li\`ege,
All\'ee du Six Ao\^ut, 17,
B-4000 Sart Tilman,
Belgium\label{inst-be}
}

   \date{}

 
  \abstract
   {One of the novelties of the \gaia-DR3 with respect to the previous data releases is the publication of the multiband light curves of about 1~million of active galactic nuclei (AGN) and of the values of some parameters characterizing their variability properties. }
   {The goal of this work was the creation of a catalogue of variable AGN, whose selection was based on \gaia data only.}
   {We first present the implementation of the methods to estimate the variability parameters into a specific object study module for AGN (SOS-AGN). Then we describe the selection procedure that led to the definition of the high-purity \gaia variable AGN sample and analyse the properties of the selected sources. We started from a sample of millions of sources, which were identified as AGN candidates by 11 different classifiers based on variability processing.
   %
    Because the focus was on the variability properties, we first defined some pre-requisites in terms of number of data points in the \gmag band and mandatory variability parameters. Then a series of filters was applied using only \gaia data and the \gaia Celestial Reference Frame~3 (\gaia-CRF3) sample as a reference.}
   {The resulting \gaia AGN variable sample, named GLEAN, contains about $872\,000$ objects, more than $21\,000$ of which are new identifications. We checked the presence of contaminants by cross-matching the selected sources with a variety of galaxies and stellar catalogues. 
   The completeness of GLEAN with respect to the variable AGN in the last Sloan Digital Sky Survey quasar catalogue is $\sim 47\%$, while that based on the variable AGN of the \gaia-CRF3 sample is $\sim 51\%$.
   The set of filters applied to the sources selected by SOS-AGN to increase the sample purity reduced the source number by about 37\%.
   From both a comparison with other AGN catalogues and an investigation of possible contaminants, we conclude that purity can be expected to be above 95\%.
   Multiwavelength properties of these sources are investigated. In particular, we estimate that $\sim 4\%$ of them are radio-loud.
   We finally explore the possibility to evaluate the time lags between the flux variations of the multiple images of strongly lensed quasars, and show one case.}
   {}


   \maketitle
%

\section{Introduction}
\label{sec:intro}

Active galactic nuclei (AGN) are present in a variety of different types, all characterized by accretion of matter onto a supermassive black hole (SMBH) with mass greater than a million solar masses. 
A fraction of AGN are radio-loud \citep[][see discussion in Sect.~\ref{sec:radio}]{jiang2007,kratzer2015} and exhibit two plasma jets that are launched from, or close to, the poles of their SMBH in opposite directions.
The members of a peculiar class of AGN, called blazars (including flat-spectrum radio quasars, FSRQs, and BL Lac-type objets), have one of the two jets closely aligned with the line of sight, which makes their multiwavelength jet emission relativistically Doppler beamed \citep{urry1995}. 

The flux of most AGN presents variability at some level, with different variability time scales and amplitudes. 
The optical continuum emission from AGN is in general dominated by the thermal radiation coming from the accretion disc, and shows smooth variability on month--year time scales. In contrast, the prevailing source of optical emission in the most active blazars is the non-thermal radiation from the relativistic jet, where Doppler beaming enhances the amplitude and reduces the time scales of variability, and even intra-night flux changes up to several tenths of magnitude can be observed \citep[e.g.][]{raiteri2017}. In objects at low redshift, the host galaxy emission can give an important contribution to, or even dominate, the optical emission, reducing the amplitude of variability. 

AGN are broadly classified in two classes. In type~1 AGN, the optical spectra show broad emission lines that are produced in a nuclear zone close to the black hole with fast-moving gas clouds. These lines are not seen in the spectra of type~2 AGN, likely because of the obscuration effect of a dusty torus. 
Narrow emission lines then appear in both type~1 and type~2 AGN spectra. They come from an outer nuclear region, where gas clouds have smaller velocities.

AGN have been identified in different ways.
A series of catalogues of spectroscopically confirmed quasars from the Sloan Digital Sky Survey\footnote{\url{https://www.sdss.org/}}(SDSS) have been published in the last about 20~years, 
from the early data release by \citet{schneider2002}, including 3814 quasars detected over $494 \, \rm deg^2$, through various releases, 
until the most recent one \citep[DR16Q,][]{lyke2020}, which contains 750\,414 quasars within $9\,376 \, \rm deg^2$. The quasar selection criteria included colour indices and variability.
The SDSS quasar catalogues have been used for a large variety of studies, from cosmology to the characterization of quasar properties.


\citet{richards2002} selected quasar candidates in the SDSS using colour indices obtained from data in the $ugriz$ filters and searching the radio counterparts of the unresolved sources in the FIRST catalogue. 
\citet{richards2009} updated the previous work by also considering the UV-excess and extended the analysis to high-redshift quasars.
A mixed selection method, including optical colours and variability, was adopted by \citet{eyer2002} and \cite{ross2012}; the latter authors used also data at other wavelengths.
Some authors proposed quasar selection methods based uniquely on variability.
\citet{macleod2011} adopted a damped random walk model to describe the temporal behaviour of quasars and to parametrize the quasar structure function. This allowed them to derive the characteristic variability time-scale and a driving amplitude of short-term variations, which are very efficient to separate quasars from stars.
Under the same assumption that the quasar temporal behaviour can be described as a damped random walk, \citet{butler2011} 
modelled the ensemble quasar structure function as a function of magnitude. This produced metrics for evaluating the probability for a source of being a quasar.

Colour indices obtained from the mid-infrared all-sky survey performed by the {\it Wide-field Infrared Survey Explorer} \citep[{\wise},][]{wright2010} satellite were found to be a superb tool to classify celestial objects, in particular AGN \citep{mateos2012,stern2012,assef2013,secrest2015,assef2018}.

Other studies have combined optical and \wise data, but were limited in sky coverage until \gaia data became available.
\citet{yan2013} used both \wise and SDSS photometry to characterize extragalactic sources and highlighted the power of \wise to identify AGN. In particular, they found that strong AGN at $z \le 3$ show $W1-W2 > 0.8$ and $W2 < 15.2$. Type-2 AGN candidates in addition require $r-W2 > 6$. 

\citet{shu2019} cross-matched the \gaia-DR2 \citep{brown2018} and unWISE \citep{schlafly2019} catalogues and used a random-forest classifier based on 16 features to select AGN. They found that the most effective features are the $W1-W2$ colours, the proper motion significance, and the extinction-corrected $G-W1$ colour. They built two catalogues: one with overall completeness of 75\% (C75), including 2\,734\,464 sources, 2\,182\,193 of which constitue a 85\% reliability catalogue (R85).

The MILLIQUAS catalogue \citep{flesch2015} contains about 2~million AGN and high-confidence candidates from  other catalogues. It has been recently updated by \citet{flesch2021}, including associations with Very Large Array Sky Survey \citep[VLASS;][]{lacy2020} radio sources.

Recently, \citet{liu2021} published a catalogue of X-ray properties of AGN in the Final Equatorial-Depth Survey (eFEDS) performed by  eROSITA\footnote{\url{https://erosita.mpe.mpg.de/edr/eROSITAObservations/Catalogues/}}.

The problem of selecting quasars at low Galactic latitudes, where extinction makes the task extremely hard, was faced by \cite{fu2021}. They built a catalogue of 160\,946 sources at $|b| \le 20 \deg$ using photometric data from Pan-STARRS1\footnote{\url{https://panstarrs.stsci.edu/}} (PS1) and AllWISE \citep{cutri2013} for classification, and \gaia proper motions to exclude stellar contaminants.

The extragalactic content of \gaia-DR2 was analysed by \citet{bailer2019}, who identified quasars and galaxies using \gaia photometric and astrometic information only. They classified 2.3 million objects as quasars, inferring that the realistic number is around 690\,000.
\citet{bailer2022} present the extragalactic content of \gaia-DR3, providing catalogues of AGN (and galaxies) that were driven by completeness, but with low purity, together with the prescriptions to obtain higher-purity samples.

The aim of the present paper is to first present the \gaia Specific Object Study package on AGN (SOS-AGN), which is part of the variability analysis pipeline discussed in \citet{eyer2022}.
The package receives inputs from the classification module \citep[see][]{rimoldini2022} and implements methods to estimate the variability characteristics of the candidate AGN.
Then, we describe the procedure that led to the selection of a high-purity sample of variable AGN and analyse its properties. Because the emphasis is on variability, among the several million AGN candidates provided by the classifiers \citep{rimoldini2022}, we consider only those sources whose \gmag light curve contains at least 20~field-of-view (FoV) transits in the \gmag~band and for which some relevant parameters can be defined. 
A set of filters is then applied, which are tailored to the properties of the 
AGN belonging to the \gaia-CRF3 sample. 
\citep{klioner2021,klioner2022}. 
Some basic information on SOS-AGN and selection results can also be found in the \gaia-DR3 online documentation\footnote{\url{https://gea.esac.esa.int/archive/documentation/GDR3}} \citep{DR3-documentation}.


An outline of the paper follows. 
In Sect.~\ref{sec:sos}, we describe the content of the SOS-AGN module, which allowed us to perform a preliminary analysis of the variability characteristics of the objects.
Section~\ref{sec:sele} details the series of cuts that we applied to remove contaminants.
The properties of the selected variable AGN sample are discussed in Sect.~\ref{sec:pro}, while a search for possible stellar contaminants is presented in Sect.~\ref{sec:con}.
Completeness and purity of our sample are addressed in Sect.~\ref{sec:copu}.
In Sect.~\ref{sec:radio}, we look for the radio counterparts of our objects and infer the fraction of radio-loud sources, while in Sect.~\ref{sec:lens} we discuss the possibility to derive time lags from the \gaia light curves of the multiple images of gravitationally lensed quasars. A brief summary and conclusions of our results are presented in Sect.~\ref{sec:fine}.

\section{SOS-AGN}
\label{sec:sos}
Our goal was to select a sample of variable AGN candidates as pure as possible.
For this, a Specific Object Study package on AGN (SOS-AGN) was implemented in the \gaia~DR3 variability pipeline \citep{eyer2022}, which depends on the upstream modules of general variability detection (GVD) and classification.
GVD pre-selected the 25\% most variable objects per magnitude interval in the \gmag~band. These variables were then classified by supervised methods. The training representatives of AGN originated mainly from \gaia-CRF3, given its high purity and all-sky distribution. The brightest known AGN were included in the training set to improve the chances of detecting rare bright AGN. All the AGN sources used for training satisfied the variability threshold of GVD. The filters applied to AGN classification results were similar to those used in SOS-AGN (as described in Sect.~\ref{sec:sele}), although with generally more permissive thresholds.

\begin{figure}
   \centering
   \includegraphics[width=\hsize]{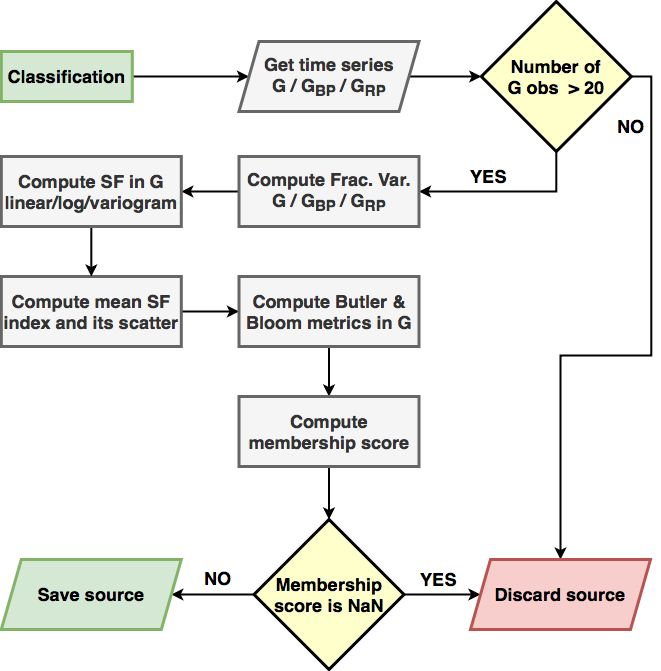}
      \caption{Flow chart of the SOS-AGN package. 
              }
         \label{fig:sos-agn}
   \end{figure}

The flow chart of the SOS-AGN processing is shown in Fig.~\ref{fig:sos-agn}. 
The first requirement for the sources to be considered was the presence of at least 20~FoV transits in the \gmag~band light curve.
Then, we defined some mandatory metrics, whose values are listed in the \gaia-DR3 \texttt{vari\_agn} table.
Mandatory means that if the parameter does not produce a real value, then the object is discarded.

The first mandatory parameter was the fractional variability \citep{vaughan2003} in the \gmag band, named \texttt{fractional\textunderscore variability\textunderscore g} in the \texttt{vari\_agn} table. 
The flux in the \gmag band was calculated as 
$F=10^{-0.4 \, (G - ZP_G)}$, where \gmag (\texttt{median\_mag\_g\_fov} in the \texttt{vari\_summary} table, here and thereafter) is the derived time series median, and $ZP_G \sim 25.7$ is the zero point in the \gmag band in the Vega system \citep{riello2021}.

To mitigate the effect of outliers, we modified the standard fractional variability definition, adopting for the flux statistics the median instead of the mean, and the median absolute deviation (MAD) instead of the standard deviation:
\begin{equation}
\texttt{fractional\_variability\_g} = 
\frac
{\sqrt{\rm MAD^2({\it F}) \, - <\sigma^2_{\it F}>}}
{\rm median({\it F})},
\end{equation}
where $<\sigma^2_{\it F}>$ is the mean of the squared flux uncertainties.
We note that because the standard deviation is approximately $1.5 \times \rm MAD$, the above definition leads to lower fractional variability values than in the classical case. In contrast, the photometric uncertainties are somewhat underestimated \citep{evans2022}, which acts in the opposite direction.

The second and third mandatory parameters were the index of the Structure Function (SF), \texttt{structure\textunderscore function\textunderscore index}, and its scatter \texttt{structure\textunderscore function\textunderscore index\textunderscore scatter}.
The slope of the SF in the $\log \rm (SF)$ versus $\log \tau$ diagram is a powerful parameter to select AGN.
There are many implementations of the SF in the literature; we adopted the classical algorithm developed by \citet{simonetti1985}.
\begin{equation}
\rm SF_{Sim} (\tau) = <[mag({\it t})-mag({\it t}+\tau)]^2>
\label{simonetti}
\end{equation}
where $\tau$ is the time lag.
The slope of the SF depends on the variability behaviour of the source and on other important physical parameters, such as redshift. AGN are known to show long-term variability, with SF slopes typically larger than 0.1 \citep[e.g.][]{eyer2002,sumi2005}.

\begin{figure}
   \centering
   \includegraphics[width=\hsize]{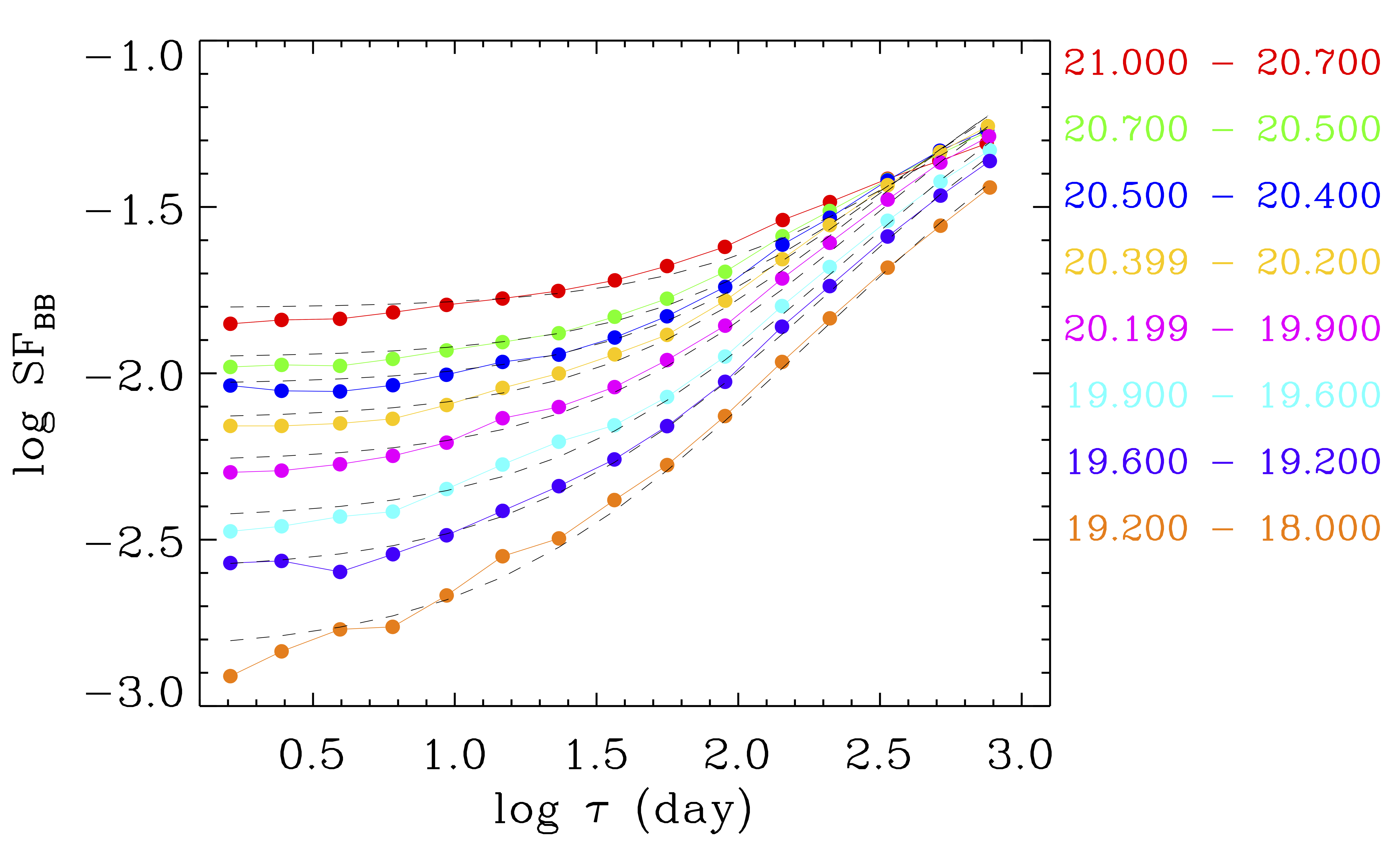}
      \caption{Mean structure functions versus time lag $\tau$ for sources with more than 5~FoV transits in the \gmag\ band in a preliminary version of the \gaia-CRF3 sample, including about 1\,850\,000 AGN candidates. The various colours correspond to different \gmag\ ranges for which the mean SFs have been estimated. Dashed lines represent the best-fit models to the mean SFs according to Eq.~\ref{eq:sf}. 
              }
         \label{fig:sf}
   \end{figure}

\begin{figure}
   \centering
    \includegraphics[width=9cm]{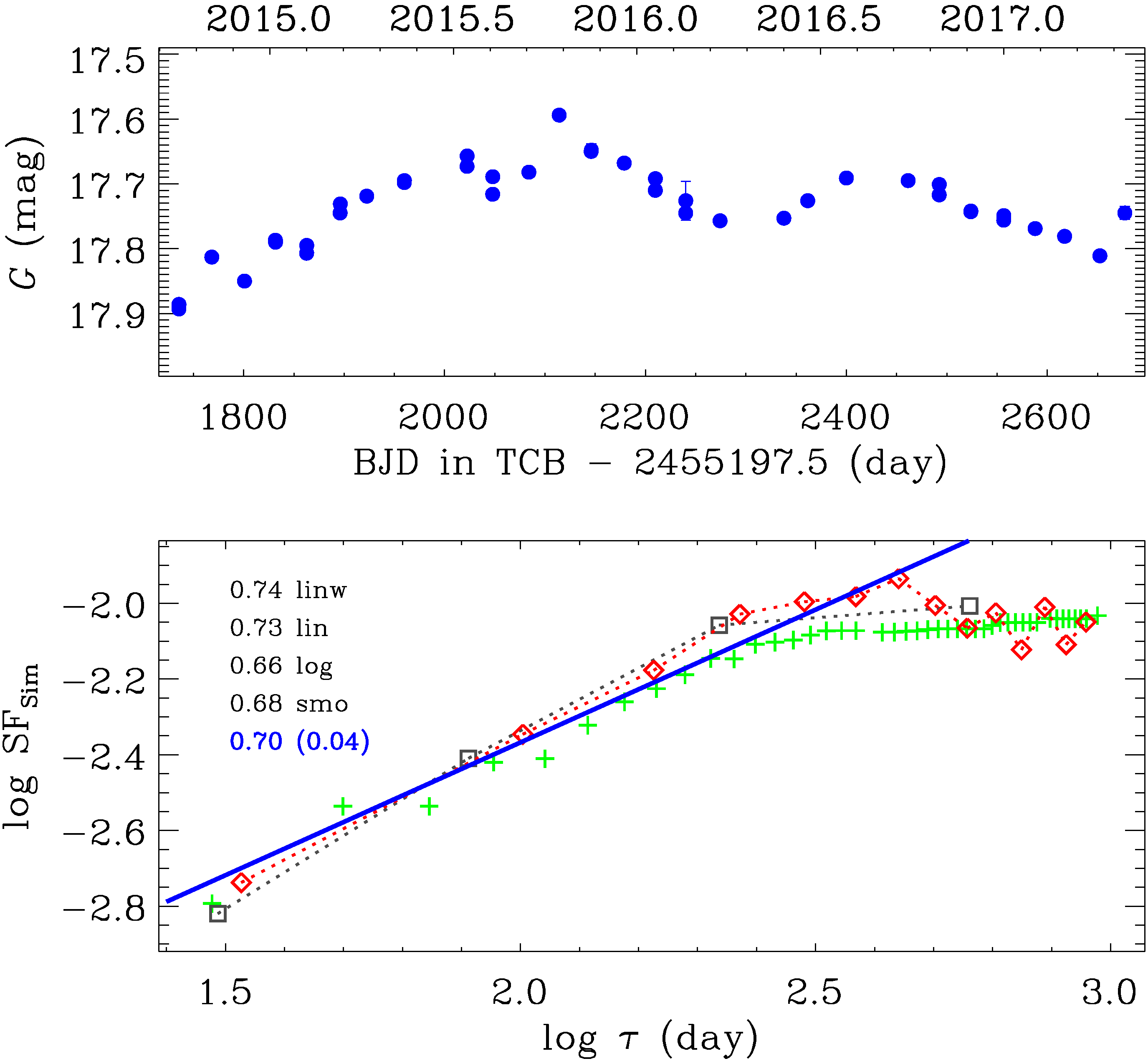}
      \caption{Top: the \gmag-band light curve of a representative variable AGN (\gaia~DR3 source\_id=4768409993534612992). Bottom: the SF obtained with linear (red diamonds) and logarithmic (grey squares) time lag sampling, and with the smoothed variogram (green plus signs). 
      The legend lists the SF slopes obtained with the four methods described in the text (black) and the final slope (blue) with its standard deviation in brackets.
      A line with this slope is drawn in blue.
              }
         \label{fig:sfes}
   \end{figure}

To implement an automatic estimate of the SF slope for every single source, we must take into account that, as $\log \tau$ increases, $\log \rm (SF)$ ideally presents first a plateau which depends on the noise, then an almost linear increase, and finally another plateau, where the second break point indicates a characteristic time-scale \citep[e.g.][]{hughes1992}. 
To calibrate the first break, we built SFs for the $\sim 1\,850\,000$ sources in a preliminary version of the \gaia-CRF3 sample, divided into magnitude bins.
Figure~\ref{fig:sf} shows the results. As the magnitude increases, the break point shifts towards longer $\tau$ values.
The second break was set where $\log \rm (SF)$ reaches its maximum value.
For each source, the SF behaviour between the two breaks was then fit with a least-square linear regression, after discarding data points with large uncertainties. Because of the dependence of the first break on magnitude mentioned above, the first break was set according to the source average magnitude, while the second break was defined by its $\log \rm (SF)$ maximum.
For each source, the linear fit was performed in four different ways, whose results were finally averaged. We considered both linear and logarithmic $\tau$ bins, and in the linear case we estimated the slope also by weighting for the number of data points in each bin. In addition to these three methods, we also estimated the slope through linear regression of a smoothed variogram \citep{eyer1999}. 
In the \gaia-DR3 \texttt{vari\_agn} table, \texttt{structure\textunderscore function\textunderscore index} represents the average value and \texttt{structure\textunderscore function\textunderscore index\textunderscore scatter} the standard deviation of these four estimates.
The bottom panel of Fig.~\ref{fig:sfes} shows an example of SF slope determination using the four methods above.

\begin{figure}
   \centering
    \includegraphics[width=9cm]{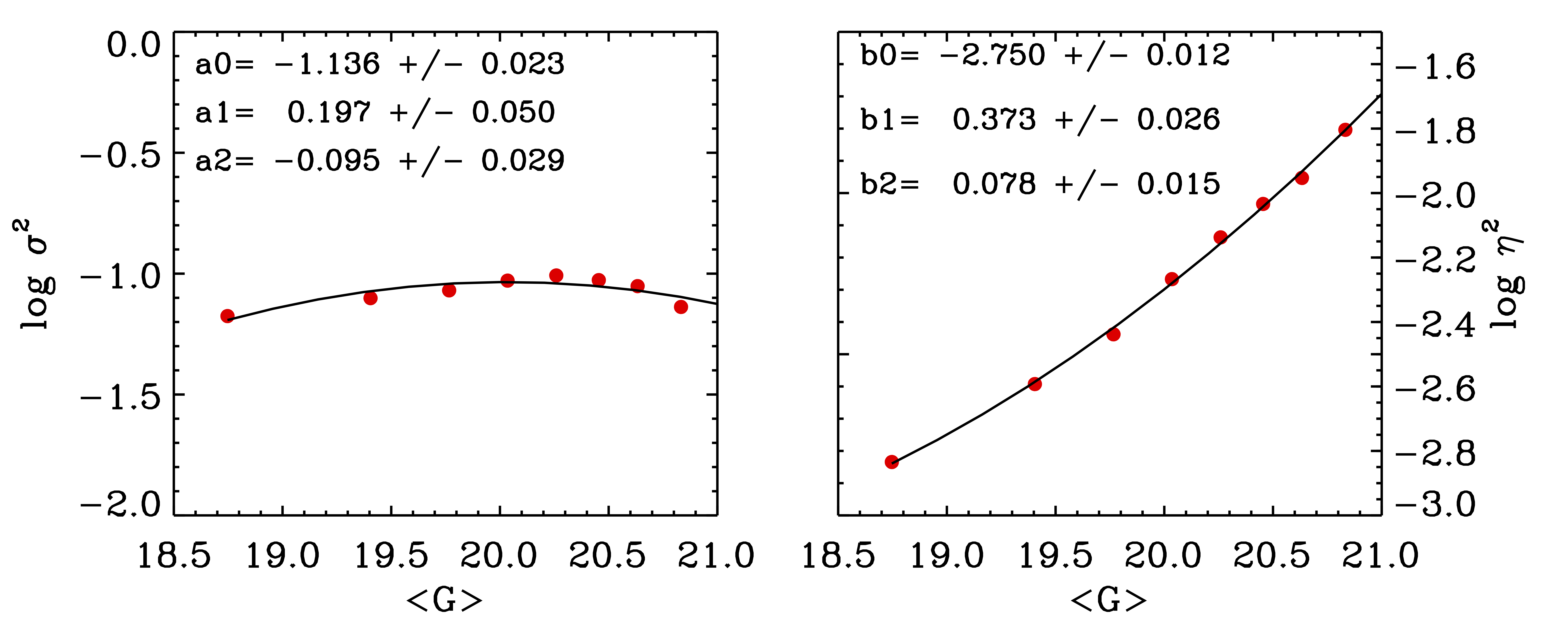}
      \caption{Results of the quadratic fits to the quantities $\log \sigma^2$ and $\log \eta^2$ defining the SF in Eq.~\ref{eq:sf}. Each data point corresponds to a mag range in Fig.~\ref{fig:sf}. The best-fit values of the parameters $a_i$ and $b_i$ in Eqs.~\ref{eq:sigma} and \ref{eq:noise} are listed in the legends.
              }
         \label{fig:bb}
   \end{figure}

The fourth and fifth mandatory parameters are the \texttt{qso\_variability} and \texttt{non\_qso\_variability} metrics (in the \texttt{vari\_agn} table) introduced by \citet{butler2011}.
As mentioned in the introduction,  \citet{butler2011} developed a method to select quasars from their light curve behaviour in a single photometric band. This is based on a damped-random walk modelling of the SF.
The parametrization of the SF made by the above authors, using the $g$-band SDSS light curves of the quasars in the Stripe~82 sky region, had to be adapted to the \gaia data. We used the expression:
\begin{equation}
\rm SF_{BB} (\tau) = \eta^2+ \sigma^2  \, [1- exp(-\tau/\tau_0)],
\label{eq:sf}
\end{equation}
where the term $\eta^2$ accounts for the noise and $\tau_0$ was fixed to 1000~days in agreement with the results of \citet{butler2011}.
The application of this model to the mean SFs in selected magnitude ranges (see Fig.~\ref{fig:sf}), whose average value is $\langle G\rangle$, allowed us to obtain the best-fit parameters $\eta^2$ and $\sigma^2$ for each magnitude bin. The trends of these parameters versus magnitude (see Fig.~\ref{fig:bb}) were then fitted by quadratic relations of the form:
\begin{equation}
\log \sigma^2=a_0+a_1 \, (\langle G\rangle-19)+a_2 \, (\langle G\rangle-19)^2
\label{eq:sigma}
\end{equation}
\begin{equation}
\log \eta^2=b_0+b_1 \, (\langle G\rangle-19)+b_2 \, (\langle G\rangle-19)^2
\label{eq:noise}
\end{equation}
to obtain the best-fit coefficients $a_i$ and $b_i$ that were required to calculate the \texttt{qso\_variability} and \texttt{non\_qso\_variability} metrics for every source\footnote{The \texttt{qso\_variability} and \texttt{non\_qso\_variability} metrics actually represent the logarithm of the corresponding \citet{butler2011} metrics.}.

\begin{figure}
   \centering
   \includegraphics[width=8cm]{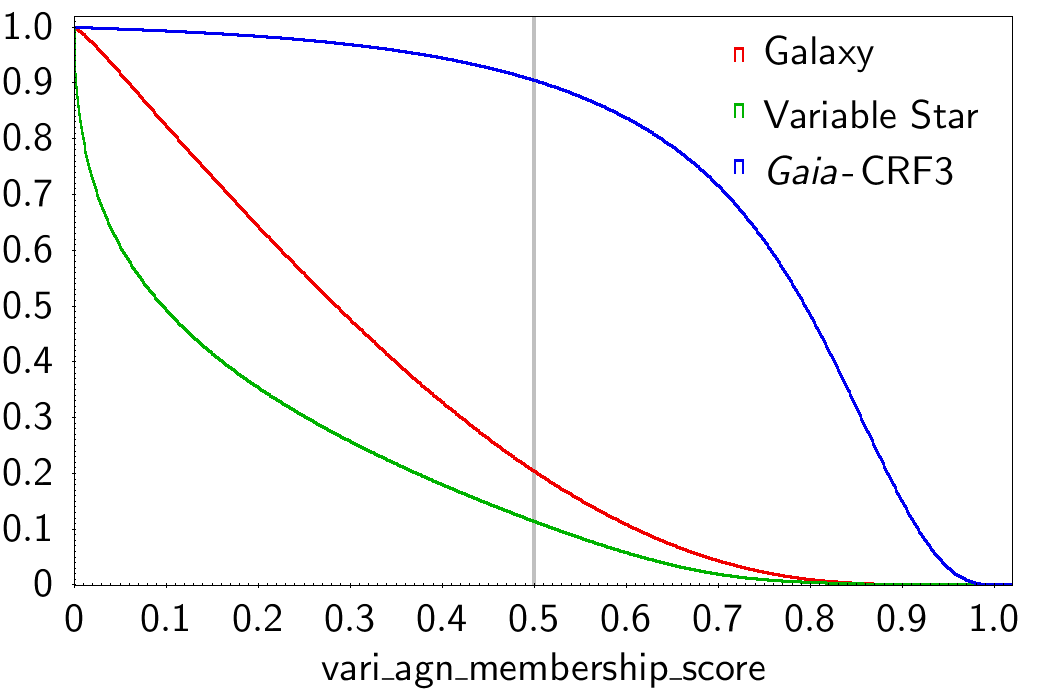}
      \caption{Normalized reverse cumulative distribution of the membership score ($1-$CDF(score)) for one million AGN from \gaia-CRF3 (blue), 0.8~million galaxies \citep[red;][]{krone2022}, and 3~million variable stars \citep[green;][]{gavras2022} from the literature. More than 90\% of the AGN have values greater than 0.5. }
         \label{fig:sig}
   \end{figure}

Finally, we defined a membership score, named \texttt{vari\textunderscore agn\textunderscore membership\textunderscore score} and published exclusively in the \texttt{qso\_candidates} table, 
which was calculated from the inverse of the Mahalanobis distance $D$ based on five parameters (\texttt{fractional\textunderscore variability\textunderscore g}, \texttt{structure\textunderscore function\textunderscore index}, \texttt{qso\textunderscore variability}, \texttt{non\_qso\textunderscore variability}, and  \texttt{abbe\textunderscore mag\textunderscore g\textunderscore fov}, where the latter is a  parameter in the \texttt{vari\_summary} table, see Sect.~\ref{sec:further}), then rescaled by a Gaussian to return values between 0 and 1:
\begin{equation}
\texttt{vari\textunderscore agn\textunderscore membershipscore}= 
 \exp[-D^2/(2 \, \rho^2)].
\label{maha}
\end{equation}
The square of the Mahalanobis distance in Eq.~\ref{maha} was computed as
$D^2=(\mathbf{x}-\mathbf{m})^T \,  C^{-1} \, (\mathbf{x}-\mathbf{m})$, where $\mathbf{x}$ is the vector of the observed values, for a given source, of the five parameters listed above, while the vector $\mathbf{m}$ of the mean values and the covariance matrix $C$ are based on the observational data for a sample of \gaia-CRF3 objects that were detected as variable by the GVD module.
The parameter $\rho$ was set to 2.7 to have more than 90\% of CRF3 sources with score larger than 0.5. Fig.~\ref{fig:sig} shows the results of a check of the \texttt{vari\textunderscore agn\textunderscore membershipscore} values on three classes of sources: AGN in the \gaia-CRF3 sample, galaxies \citep{krone2022}, and variable stars \citep{gavras2022}. As can be seen, the distribution of scores for \gaia-CRF3 sources is distinct from that of the other two classes of objects.

\section{Selection procedure}
\label{sec:sele}

Samples of variable AGN candidates with corresponding probabilities were provided by 11 classifiers in the \gaia DR3 variability pipeline, based on different prescriptions. For details, see \citet{rimoldini2022}. 
About 34~million sources from different classifiers met the criteria defined by the SOS-AGN module. 
However, to 
reduce the sample to the most reliable candidates, 
for each classifier we compared the probabilities of the AGN candidates with those of the \gaia-CRF3 sources therein and set minimum probability thresholds so that no more than 5\% of the CRF3 sources was lost.
This resulted in a sample of 10~million sources with more than 20~FoV transits in their \gmag band light curves, and which have the mandatory parameters described in Sect.~\ref{sec:sos} defined.
Among them, 1.1~million are included in the \gaia-CRF3 sample.

The selection procedure continued with the application of a sequence of filters tailored on the \gaia-CRF3 sources. The goal was to obtain a variable AGN sample as pure as possible, with the minimum loss of \gaia-CRF3 objects.
In the following, we describe the subsequent filters which were adopted to remove contaminants.
We stress that the same names are used  to denote both the initial sample and the subsamples that are derived from it as a result of the various steps in the selection procedure.
As an example, the term `\gaia-CRF3' indicates both the original sample and the various ensembles of sources belonging to it that survive the subsequent selection cuts. 

\subsection{Structure Function Index}
Following the considerations in Sect.~\ref{sec:sos}, we decided to keep candidates that satisfied the condition

\begin{figure}
   \centering     
   \includegraphics[width=8cm]{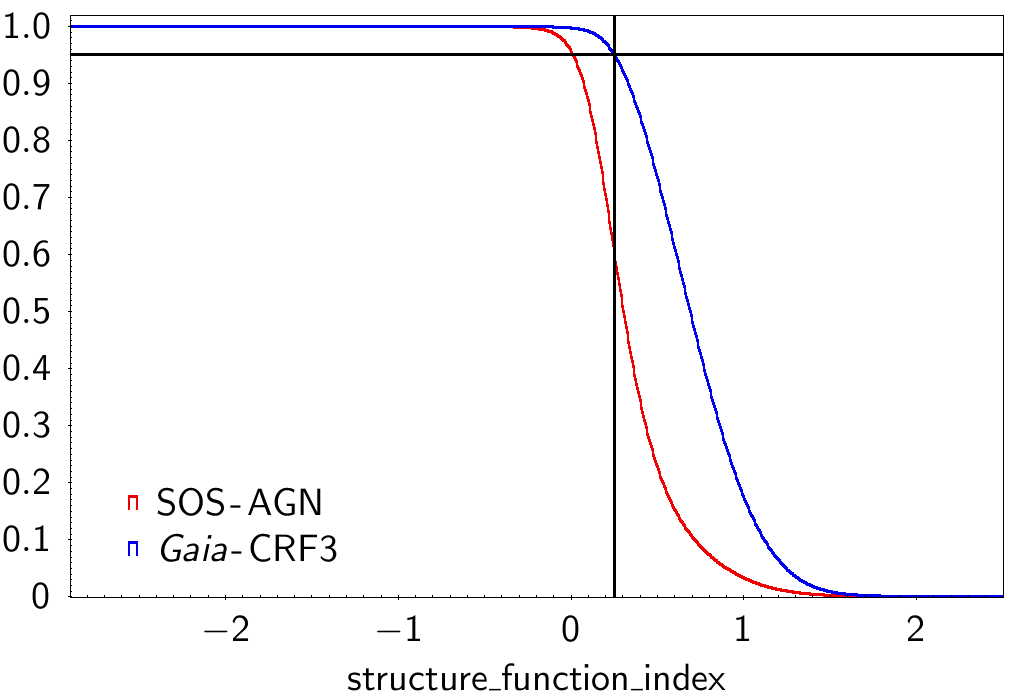}
      \caption{Normalized reverse cumulative distribution of the \texttt{structure\_function\_index} for the $\sim 10$ million variable AGN candidates (red) and for the \gaia-CRF3 sources (blue). The vertical line indicates the threshold of 0.25, i.e., the minimum value of the index required to pass the selection.
              }
         \label{fig:slope_cut}
   \end{figure}

\texttt{structure\_function\_index} > 0.25.

\noindent
where the \texttt{structure\_function\_index} is the slope of $\rm \log SF_{Sim}$  (see Eq.~\ref{simonetti}) versus $\log \tau$.
As it is shown in Fig.~\ref{fig:slope_cut}, in this way we lost about 40\% of dubious variable AGN candidates, but only 5\% of \gaia-CRF3 sources, remaining with 1~million  \gaia-CRF3 objects and 6.2~million candidates.

\subsection{QSO versus non-QSO statistics}

\begin{figure}
   \centering
   \includegraphics[width=8cm]{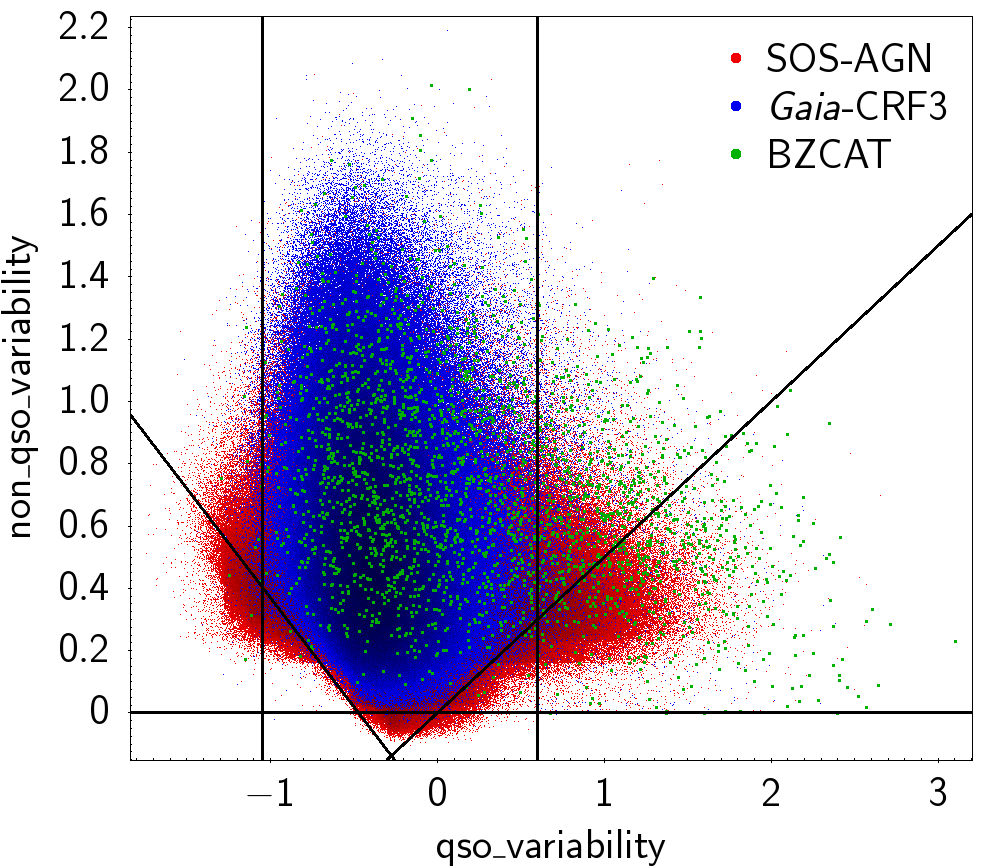}
      \caption{\citet{butler2011} metrics (actually, their logarithm) \texttt{non\_qso\textunderscore variability} versus \texttt{qso\textunderscore variability} plot, showing the position of the variable AGN candidates (red dots), distinguishing those in the \gaia-CRF3 sample (blue dots), and the blazars in the BZCAT5 catalogue (green dots). The lines highlight the cuts performed to remove contaminants.
              }
         \label{fig:bb_cut}
   \end{figure}

The \citet{butler2011} metrics were used for further cuts, defined by the region in the \texttt{qso\_variability} versus \texttt{non\_qso\_variability} space expected to host the vast majority of AGN.
The \gaia-CRF3 sources confirmed the locations of \texttt{qso\_variability} around zero and of \texttt{non\_qso\_variability} above zero. 
We defined the following cuts, where some margin was left to minimise the loss of bona fide CRF3 sources (see Fig.~\ref{fig:bb_cut}):

$\texttt{qso\_variability} > -1.05$

$\texttt{qso\_variability} < 0.6$

$\texttt{non\_qso\_variability} > 0$

$\texttt{non\_qso\_variability} > -0.7 \times \texttt{qso\_variability}-0.33$

$\texttt{non\_qso\_variability} >0.5 \times \texttt{qso\_variability}$.

\noindent 
In Fig.~\ref{fig:bb_cut}, we highlight the sources included in the fifth edition of the Roma-BZCAT blazar catalogue \citep[BZCAT5;][]{massaro2015}. Their distribution in \texttt{qso\_variability} is wider than the one of typical AGN.

\noindent
After the above selections, we were left with $\sim$6~million candidates, while the number of \gaia-CRF3 sources remained around 1~million. This filter removed many blazars (about 35\%), whose \texttt{qso\_variability} values extended to quite larger values than those of AGN in the \gaia-CRF3.

\subsection{Further filtering}
\label{sec:further}

\begin{figure}
   \centering
   \includegraphics[width=8cm]{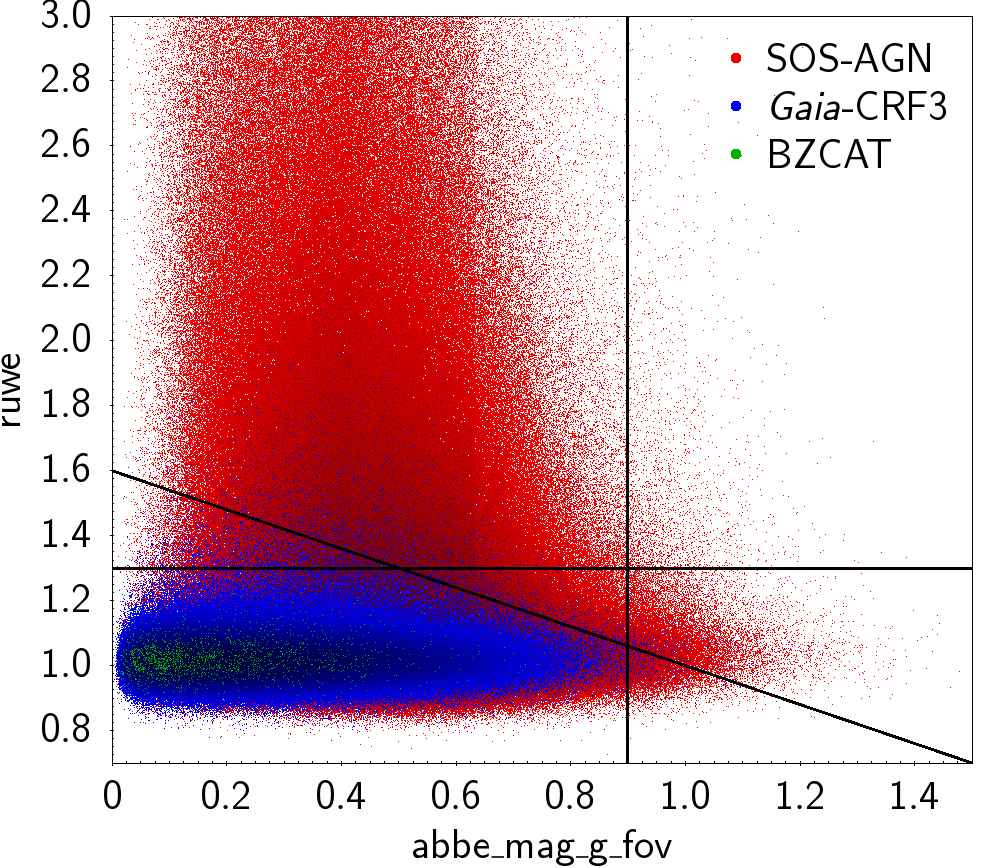}
      \caption{As Fig.~\ref{fig:bb_cut}, but for
      \texttt{ruwe} versus \texttt{abbe\_mag\_g\_fov}.
      }
         \label{fig:abbe_cut}
   \end{figure}

Constraints were set on the 
\texttt{abbe\_mag\_g\_fov} (from the \texttt{vari\_summary} table) and renormalised unit weight error (\texttt{ruwe}, in the \texttt{gaia\_source} table) parameters.
The  \texttt{abbe\_mag\_g\_fov} is defined as half of the ratio of the mean square difference between consecutive data points in the \gmag band light curve to its variance (small values correspond to time series that are smooth  in time). The \texttt{ruwe} parameter gives an estimate of the suitability of the single-star astrometric model for a given source (values close to one indicate a good agreement).
The AGN light curves generally exhibit long term variations, which are often sufficiently resolved by \gaia's sampling to cause a tendency towards small values of \texttt{abbe\_mag\_g\_fov}. 
Moreover, most AGN appear as astrometrically stable point sources, hence they are usually associated with \texttt{ruwe} values close to one. 
The \gaia-CRF3 sources confirm such expectations as they populate a compact strip in the \texttt{ruwe} versus \texttt{abbe\textunderscore mag\textunderscore g\textunderscore fov} space.
Thus, the selection region was defined as follows (see Fig.~\ref{fig:abbe_cut}):

$\texttt{ruwe} < 1.3$

$\texttt{ruwe} < -0.6 \times \texttt{abbe\_mag\_g\_fov}+1.6$

$\texttt{abbe\_mag\_g\_fov} < 0.9$.

\noindent
These 2D-cuts decreased the SOS-AGN sample to around 4.8~million of AGN candidates (still $\sim$1~million in \gaia-CRF3). This filter has only a minor effect on blazars, reducing them by about 1.4\%.

\begin{figure}
   \centering
   \includegraphics[width=8cm]{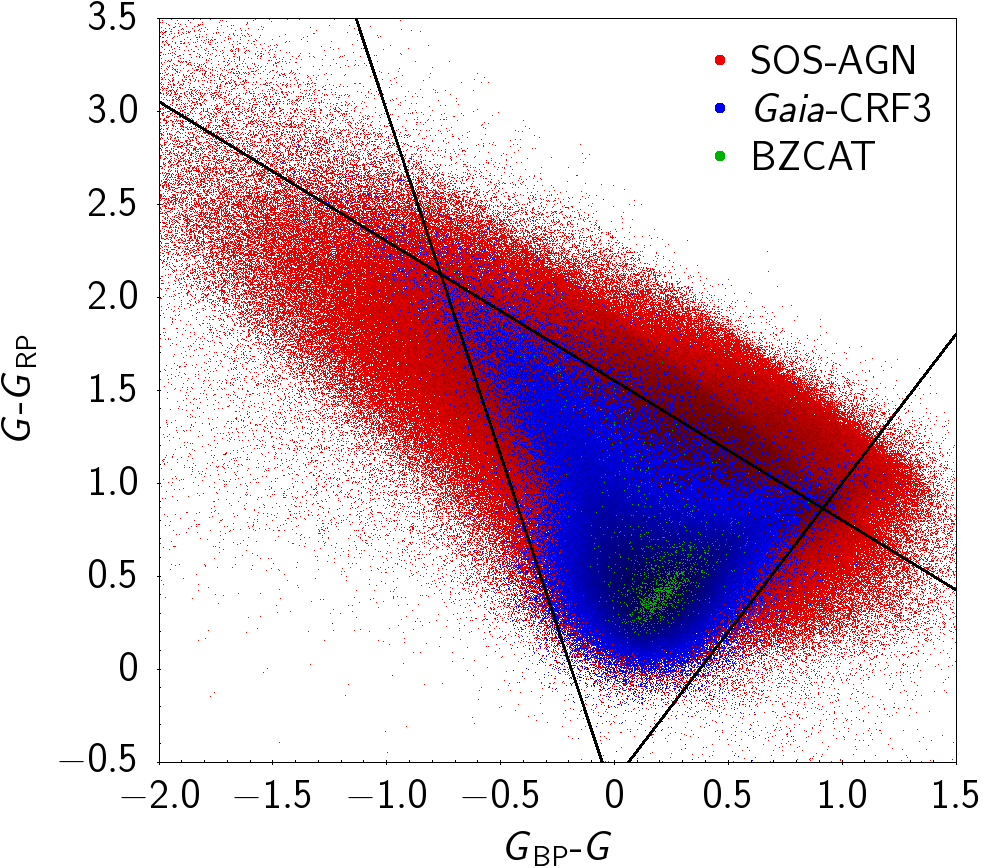}
      \caption{As Fig.~\ref{fig:bb_cut}, but for $G-G_{\rm RP}$ versus $G_{\rm BP}-G$.
              }
         \label{fig:color_cut}
   \end{figure}

Optical colour indices have proved to be important for quasar selection, even if not decisive in general.
We used \gaia colours derived from time series medians, which are found in the \texttt{vari\_summary} table.
We filtered the sources in the \bmg (\texttt{median\_mag\_bp}$\,-\,$\texttt{median\_mag\_g\_fov}) versus \gmr (\texttt{median\_mag\_g\_fov}$\,-\,$\texttt{median\_mag\_rp}) region enclosed within the following conditions (see Fig.~\ref{fig:color_cut}):

$G-G_{\rm RP} > -3.7 \times G_{\rm BP}-G-0.7$ 

$G-G_{\rm RP} < -0.75 \times G_{\rm BP}-G+1.55$

$G-G_{\rm RP} > 1.6 \times G_{\rm BP}-G-0.6$.

\noindent
This led to around 3.9~million candidates (still $\sim$1~million in \gaia-CRF3).

Extragalactic sources should ideally have null (statistically insignificant) parallax and proper motions. Therefore, these astrometric parameters (included in the \texttt{gaia\_source} table) can efficiently help to remove Galactic contaminants. 
To take uncertainties into account, the corresponding cut was made on the ratio of these parameters to their errors. Such ratios are expected to follow a normal distribution with unit variance and zero mean. A permissive condition kept candidates within 5-sigma \citep[see also][]{klioner2022}. For parallax:
\begin{equation}
|(\texttt{parallax}+0.017)/\texttt{parallax\textunderscore error}| < 5,
\end{equation}
where the addition of $\rm 0.017 \, mas$  to \texttt{parallax} takes into account the global parallax zero point of \gaia~EDR3 \citep{lindegren2021a,lindegren2021b}.
For proper motion (\texttt{pm}):
 \begin{equation*}
      \texttt{pm} = \sqrt{\frac{\alpha^2 + \beta^2 - 2 \, \alpha \, \beta \, \gamma}{1- \gamma^2}} <5,
\end{equation*}
where the \texttt{pm} components along the Equatorial coordinates, their uncertainties and correlation are taken into account as follows: 
$\alpha=\texttt{pmra}/\texttt{pmra\_error}$, 
$\beta=\texttt{pmdec}/\texttt{pmdec\_error}$, and
$\gamma=\texttt{pmra\_pmdec\_corr}$.
After this selection, the number of SOS-AGN candidates was 1.6~million. 

To reduce AGN misclassification in crowded stellar fields, e.g., in the Galactic Plane and Magellanic Clouds, we set a constraint on the environment of each candidate, limiting the maximum number density of sources within 100~arcsec to 0.004~arcsec$^{-2}$.
About 1.2~million sources passed this requirement. 

Artificial variability is produced by the scan angle 
variations for extended objects
\citep[see][]{holl2022}, as it may happen for detectable AGN host galaxies. 

We then set an upper limit to the Spearman correlation between the \gmag-band time series and the model of the Image Parameter Determination (IPD) $r_\text{ipd}$ (at scan angles corresponding to the time series observations), which quantifies the amount of scan-angle dependent signal in the photometric time series \citep[see][for details]{holl2022}.
The constraint 
$r_\text{ipd} < 0.8$
removed only about 2000 sources.

A final cut on the GVD variability probability was also made to further increase the sample purity, in view that part of the variations might be due to a spurious signal when the host galaxy is detectable.
The final list of variable AGN candidates contains 872\,228 sources, 150\,017 of which are not included in \gaia-CRF3.
It also contains almost 3\,000 objects that did not pass the selection procedure because of peculiar properties (like blazars, lensed AGN, and the brightest known AGN), but were added to the final sample for their interest.

In the rest of this article, we will refer to the whole set of selected \gaia variable AGN as the GLEAN (Gaia variabLE AgN) sample, and to those objects that are in the GLEAN sample but not in the \gaia-CRF3 one as the CANOE (CANdidates tO Explore) sample.
The CANOE objects represent an interesting AGN candidate subsample to be explored in view of a possible future addition to the \gaia-CRF3.

\section{The \gaia variable AGN sample}
\label{sec:pro}

\begin{figure}
   \centering
   \includegraphics[width=\hsize]{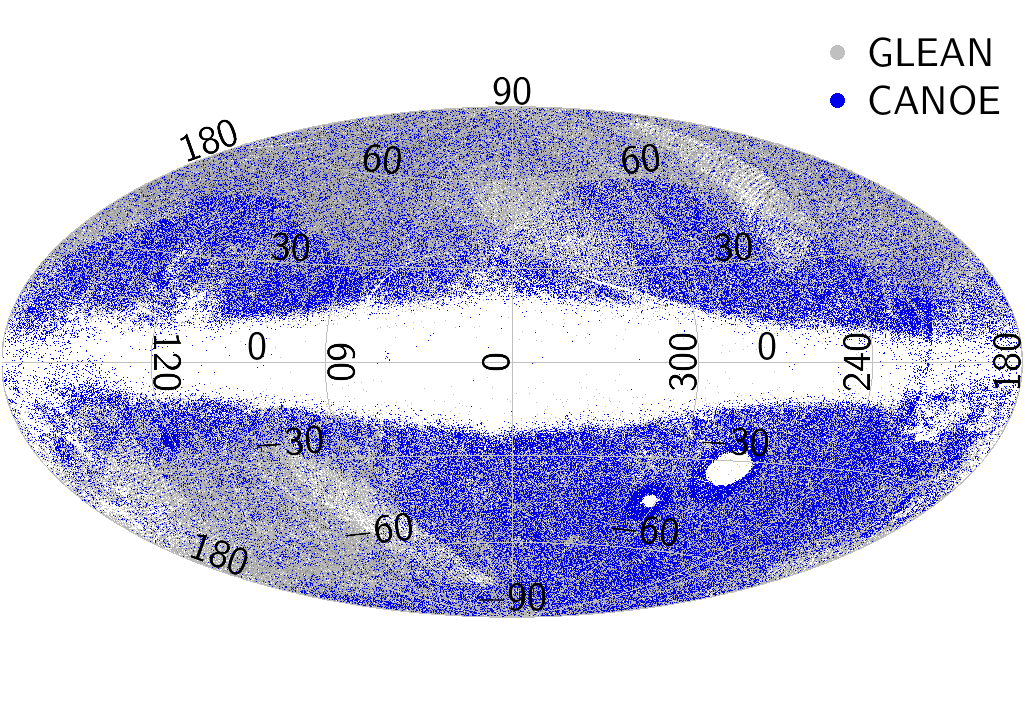}
      \caption{Sky distribution of the sources in the GLEAN (grey) and CANOE (blue) samples in Galactic coordinates. Mostly because of the environment filter, the Galactic Plane and Magellanic Clouds are almost empty. Some scanning law footprints are still visible
              }
         \label{fig:skymap_final}
   \end{figure}

The sky distribution of the sources in the GLEAN sample is shown in Fig.~\ref{fig:skymap_final}. The Galactic Plane and Magellanic Clouds are almost empty, as expected because of the filters applied, in particular that on the environment.
However, there is still an excess of AGN around the Magellanic Clouds, which may indicate some stellar contamination, or that these regions are still partially unexplored from an extragalactic point of view.
   
\begin{figure}
   \centering
   \includegraphics[width=8cm]{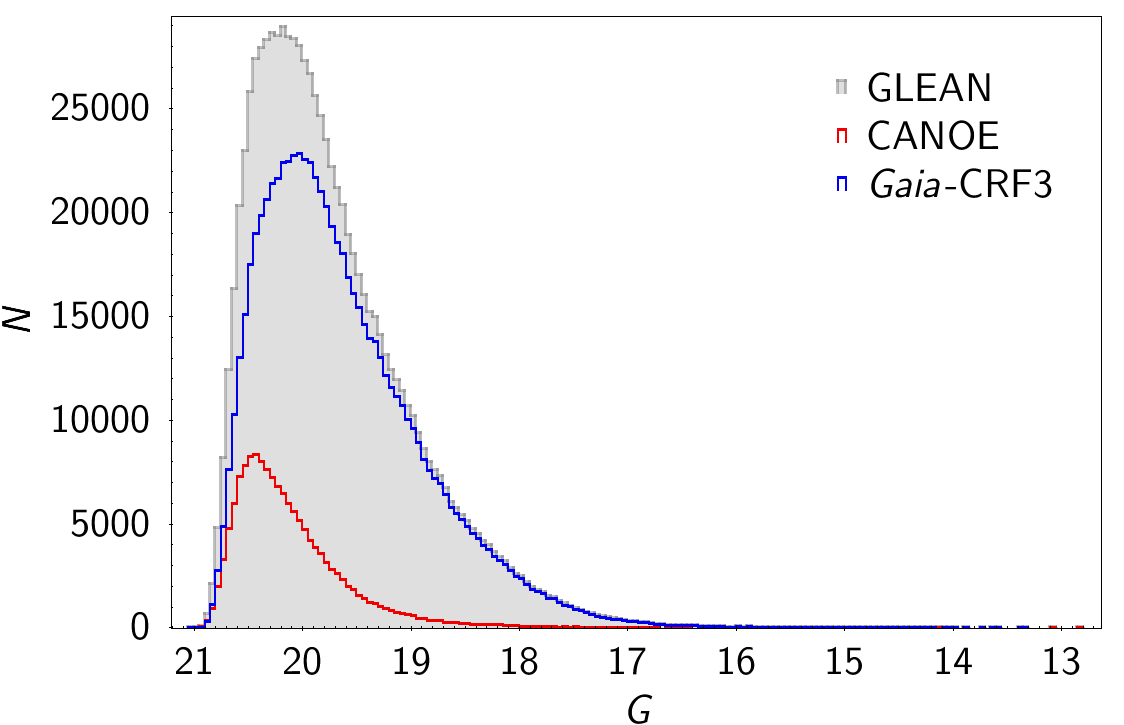}
      \caption{Magnitude distribution (\texttt{median\_mag\_g\_fov}) of all sources in the GLEAN (grey), CANOE (red), and \gaia-CRF3 (blue) samples, in bins of 0.05 mag.}
         \label{fig:gmag}
   \end{figure}

   The \gmag magnitude distribution (\texttt{median\_mag\_g\_fov}) is plotted in Fig.~\ref{fig:gmag} for the complete GLEAN sample, and for those sources in the sample that belong to the CANOE and \gaia-CRF3 sub-samples. The distribution of the CANOE sources peaks at a fainter magnitude than that of the CRF3 objects. 
   
      \begin{figure}
   \centering
   \includegraphics[width=\hsize]{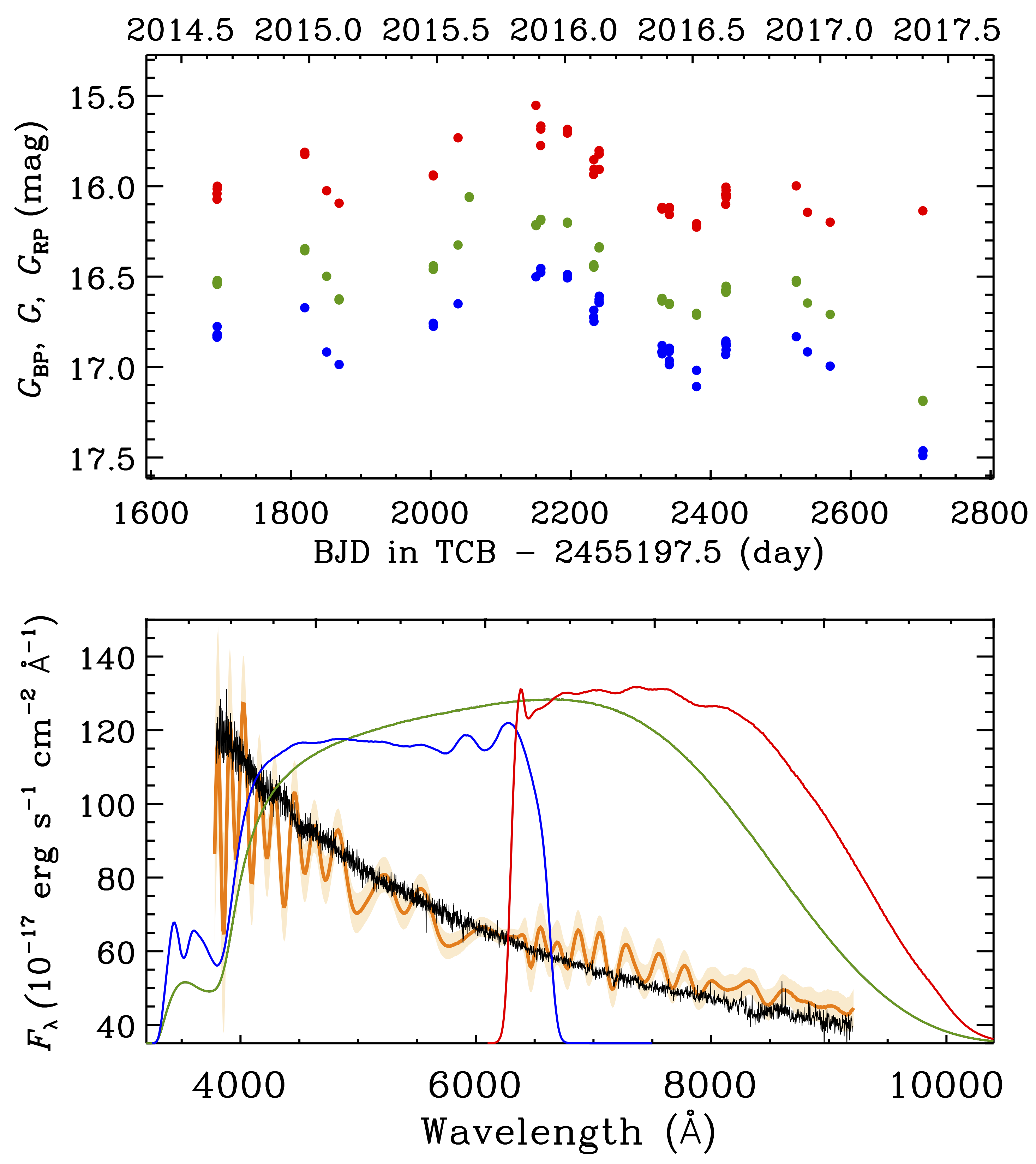}
      \caption{Top: \gmag (green), \grp (red), and \gbp (blue) light curves of the BL Lac-type source 5BZBJ0035+1515 (\gaia~DR3 source\_id: 2780475069095852672). Bottom: SDSS spectrum (black), \gaia low-resolution spectrum (orange) with its uncertainty (shadowed orange region), and \gaia passbands. 
      }
         \label{fig:bllac}
   \end{figure}

   \begin{figure}
   \centering
   \includegraphics[width=\hsize]{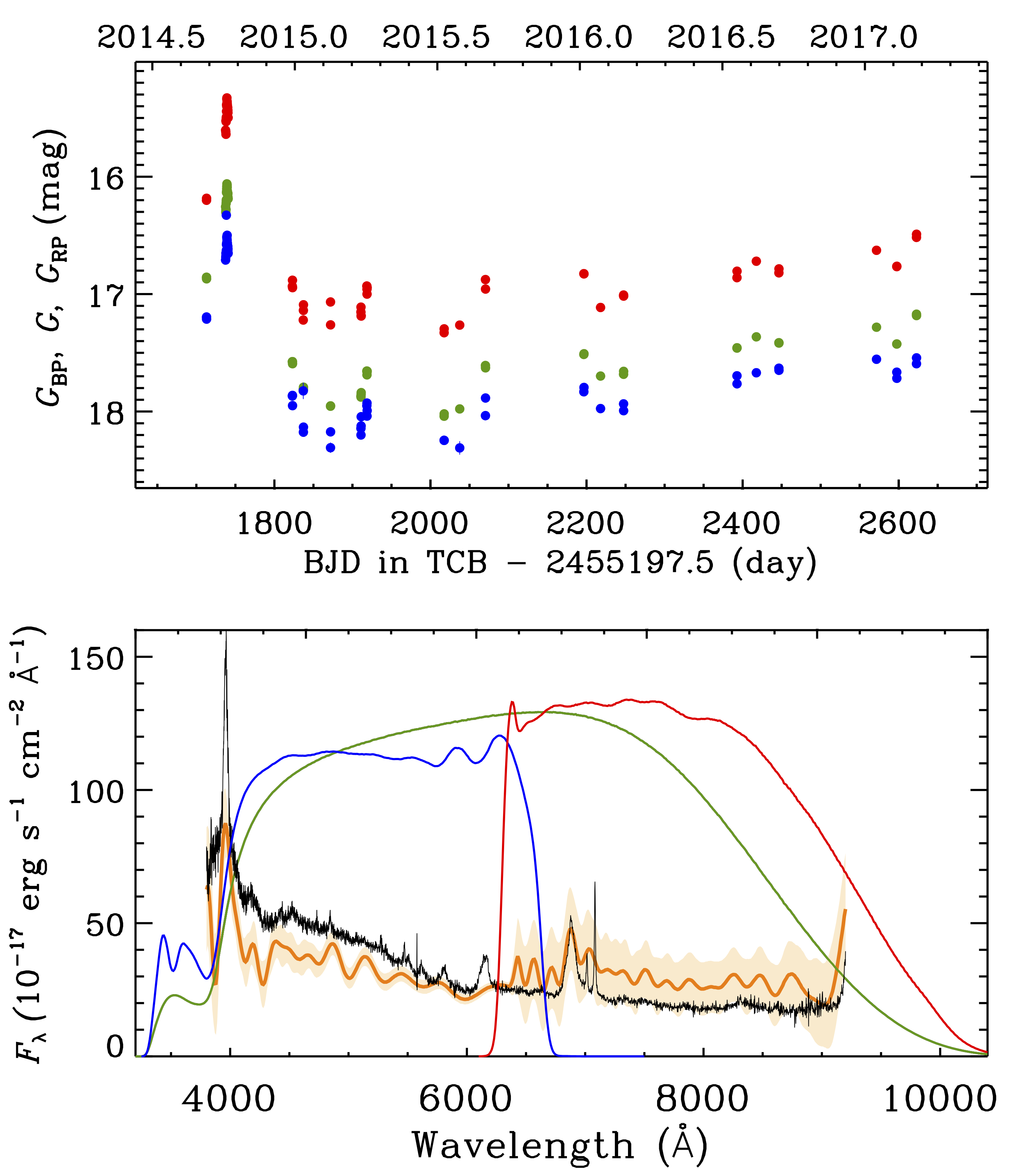}
      \caption{As Fig.~\ref{fig:bllac}, but for the FSRQ 5BZQJ1549+0237 (\gaia~DR3 source\_id: 4423448219003043968).
              }
         \label{fig:fsrq}
   \end{figure}

\begin{figure}
   \centering
   \includegraphics[width=\hsize]{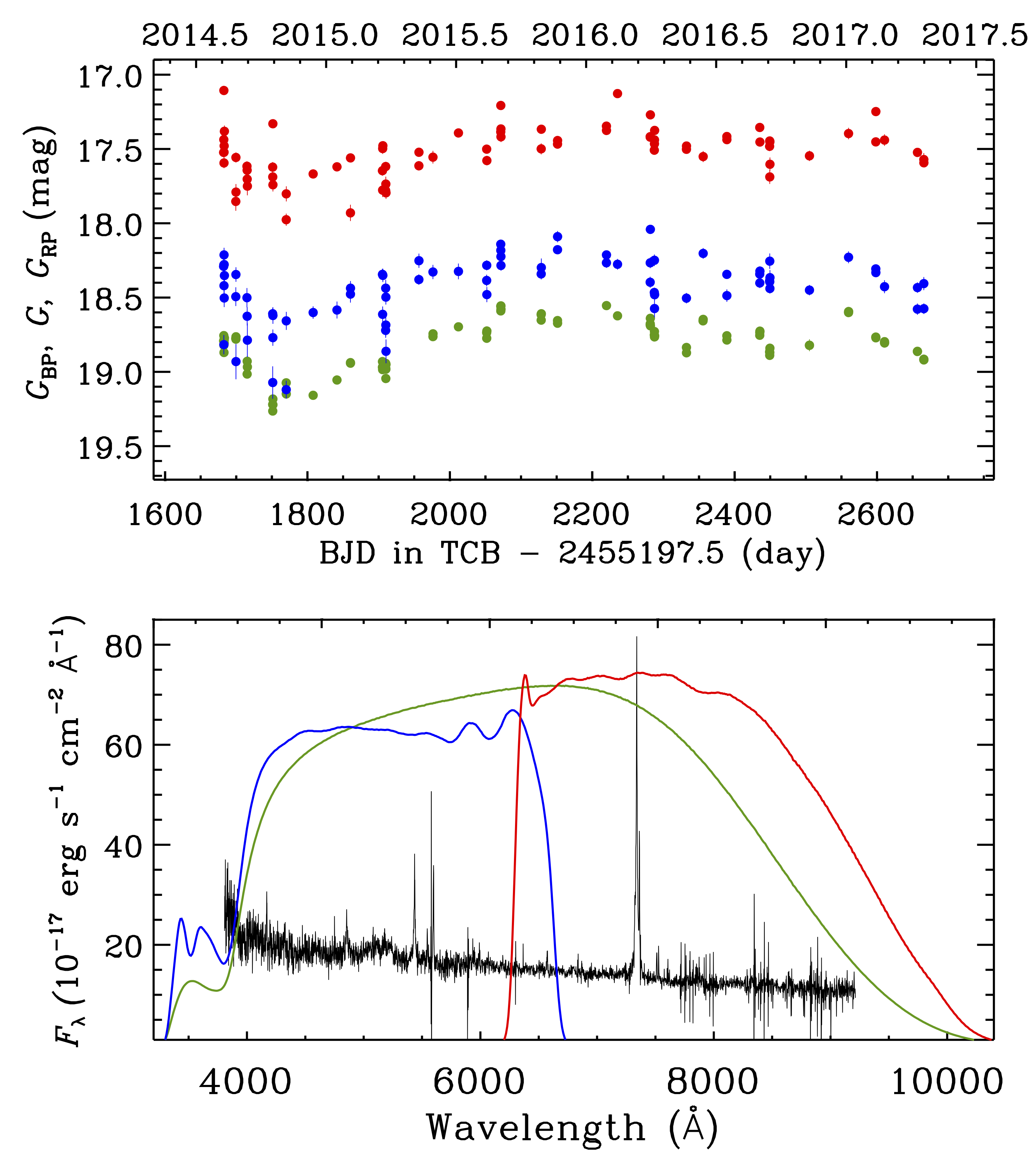}
      \caption{As Fig.~\ref{fig:bllac}, but for the Narrow Line Seyfert~1 galaxy WISEA J133928.49+403229.9 (\gaia~DR3 source\_id: 1500096699133497600).
              }
         \label{fig:seyfert}
   \end{figure}

\begin{figure}
   \centering
   \includegraphics[width=\hsize]{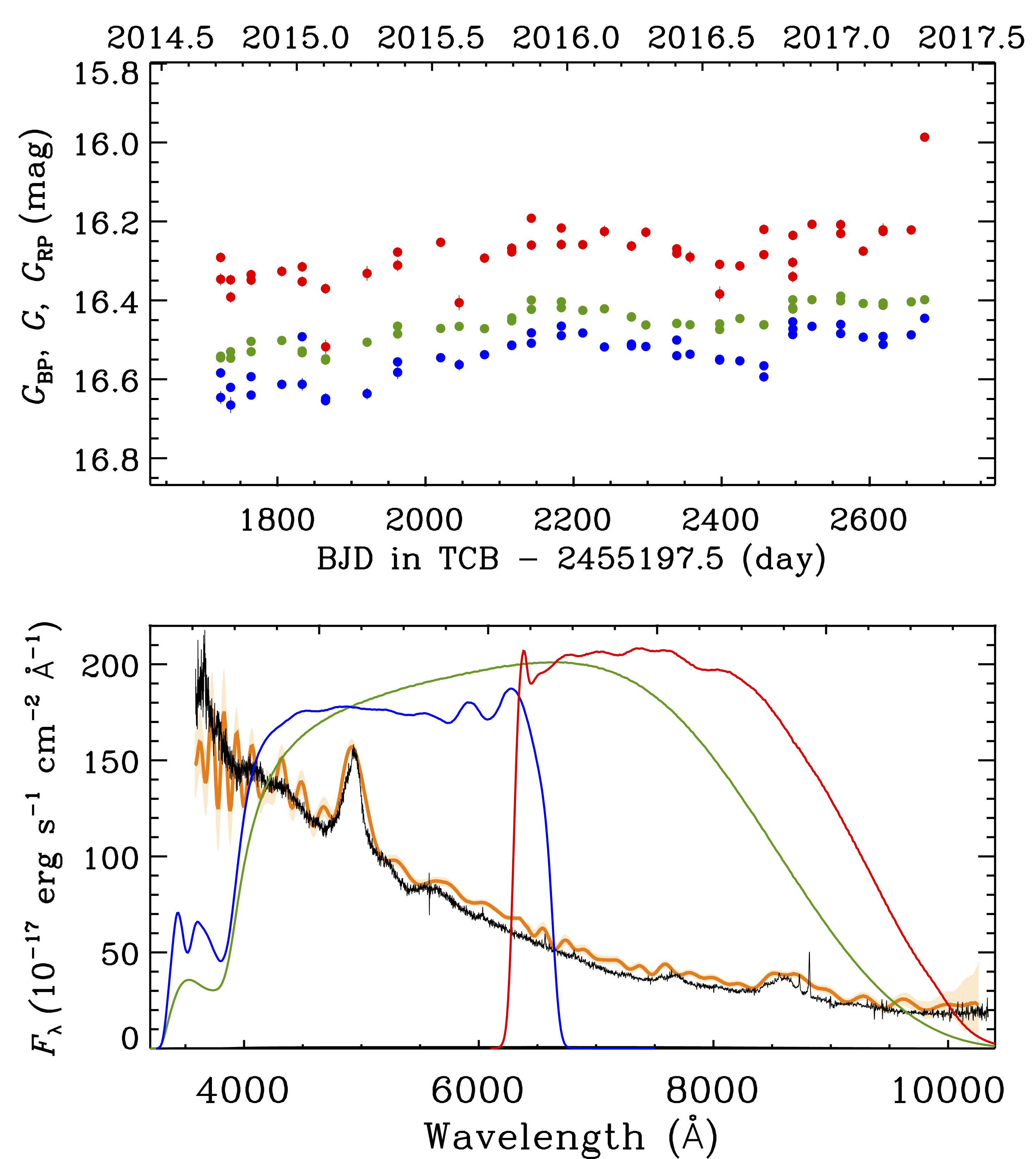}
      \caption{As Fig.~\ref{fig:bllac}, but for the radio-quiet quasar FBQS J163709.3+414030 (\gaia~DR3 source\_id: 1356927713819217664).
              }
         \label{fig:qso}
   \end{figure}

  One of the main novelties of \gaia~DR3 is the publication of the light curves for the AGN selected in this paper and in the paper by \citet{rimoldini2022}.
  Figures~\ref{fig:bllac}--\ref{fig:qso} display the \gaia multiband light curves of four representative sources: a BL Lac-type object, a flat-spectrum radio quasar (FSRQ), a Seyfert galaxy, and a radio-quiet quasar. The figures also show the SDSS spectra \citep{abolfathi2018} and the \gaia passbands, to highlight the spectral coverage of the \gaia filters.
  For three of these sources, \gaia low-resolution spectra are available in DR3 and are shown in the same figures. Details on their calibration can be found in \citet{carrasco2021}, \citet{deangeli2022} and \citet{montegriffo2022}. We underline that \gaia spectroscopic information has not been used in the variable AGN candidates selection performed in this paper.
  
  The light curves of the BL Lac-type object (Fig.~\ref{fig:bllac}) show more than 1~mag variability in the \gmag band; the SDSS spectrum is featureless, confirming that the dominant contribution is synchrotron emission from the jet. A rapid flare characterizes the light curve of the FSRQ (Fig.~\ref{fig:fsrq}) at the beginning of the \gaia monitoring, with a brightness decrease of about 2~mag, followed by a slow brightness increase. The SDSS spectrum includes the main emission lines usually present in quasar spectra, redshifted to $z \sim 0.414$. This indicates a strong emission contribution from the broad line region, in addition to that of the jet. The light curves of the Seyfert galaxy (Fig.~\ref{fig:seyfert}) show smooth variability, with maximum amplitude of about 0.7~mag in the \gmag~band and some dispersion of the data points acquired in the same Julian day, especially in the \gbp band. Because of the source faintness, the SDSS spectrum is somewhat noisy, but clearly shows the typical features of a Narrow Line Seyfert~1 galaxy redshifted to $z \sim 0.118$.
  Smooth variability (with some noise) characterizes also the light curves of the radio-quiet quasar (Fig.~\ref{fig:qso}); the SDSS spectrum shows emission lines, in particular a prominent broad  Mg II $\lambda \lambda 2796, 2803$, redshifted to $z \sim 0.761$.

\begin{figure}
   \centering
   \includegraphics[width=\hsize]{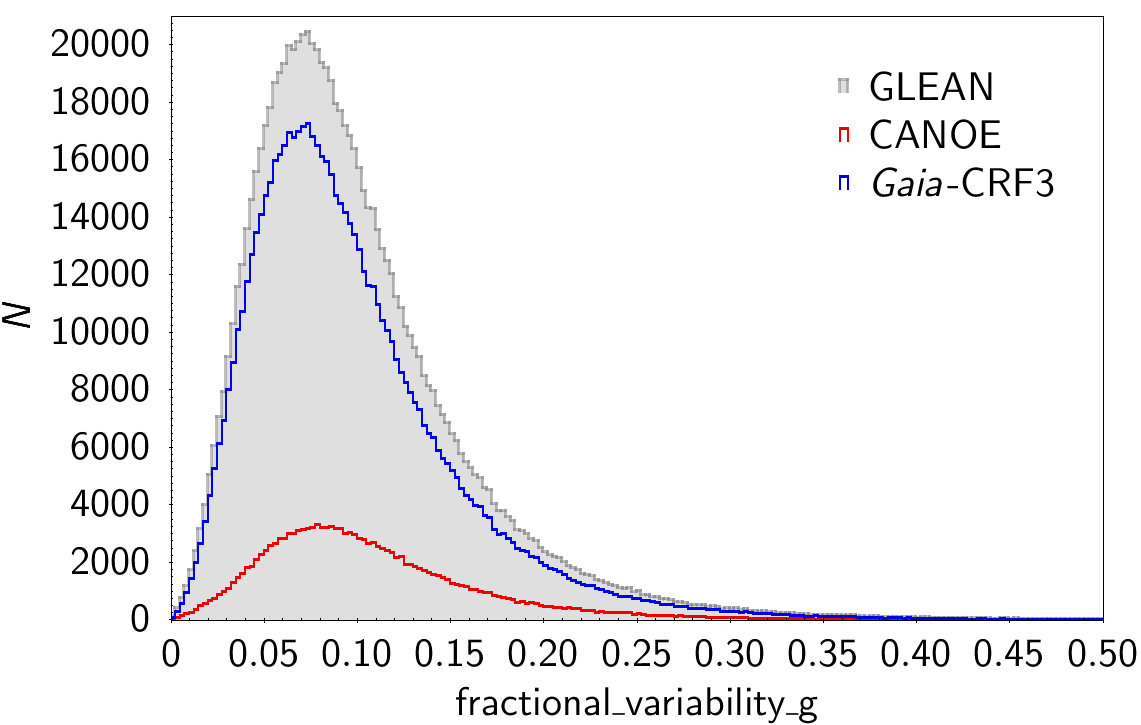}
      \caption{Distribution of the \texttt{fractional\_variability\_g} parameter for the GLEAN, CANOE, and \gaia-CRF3 samples (bin width=0.0025). The peaks indicate variability at a 7--8\% level.}
         \label{fig:fvar}
   \end{figure}

   The amount of variability can be described by the \texttt{fractional\_variability\_g} parameter (see Sect.~\ref{sec:sos}).
   Figure~\ref{fig:fvar} shows its distribution for the GLEAN, CANOE, and \gaia-CRF3 sources: the peak value for the three samples is similar, and indicates variability at 7--8\% level. Only a minority of objects has values larger than 20\%. 
   These results seem in agreement with those obtained by \citet{berghea2021}, who analysed the optical variability properties of 2863 sources belonging to the radio International Celestial Reference Frame 3 (ICRF3) with Pan-STARRS DR2 data. They found that the distributions of variability amplitudes is strongly skewed towards small values and peaks at about 0.1 mag.
   
   
We searched for infrared counterparts of our candidates in the AllWISE archive\footnote{\url{https://irsa.ipac.caltech.edu/Missions/wise.html}}.
We considered a 3~arcsec search radius and asked for a signal-to-noise (SNR) greater than 3 in the $W1$, $W2$, and $W3$ bands, obtaining 569\,530 matches (53\,144 of which are CANOE sources).
%

\begin{figure}
   \centering
   \includegraphics[width=\hsize]{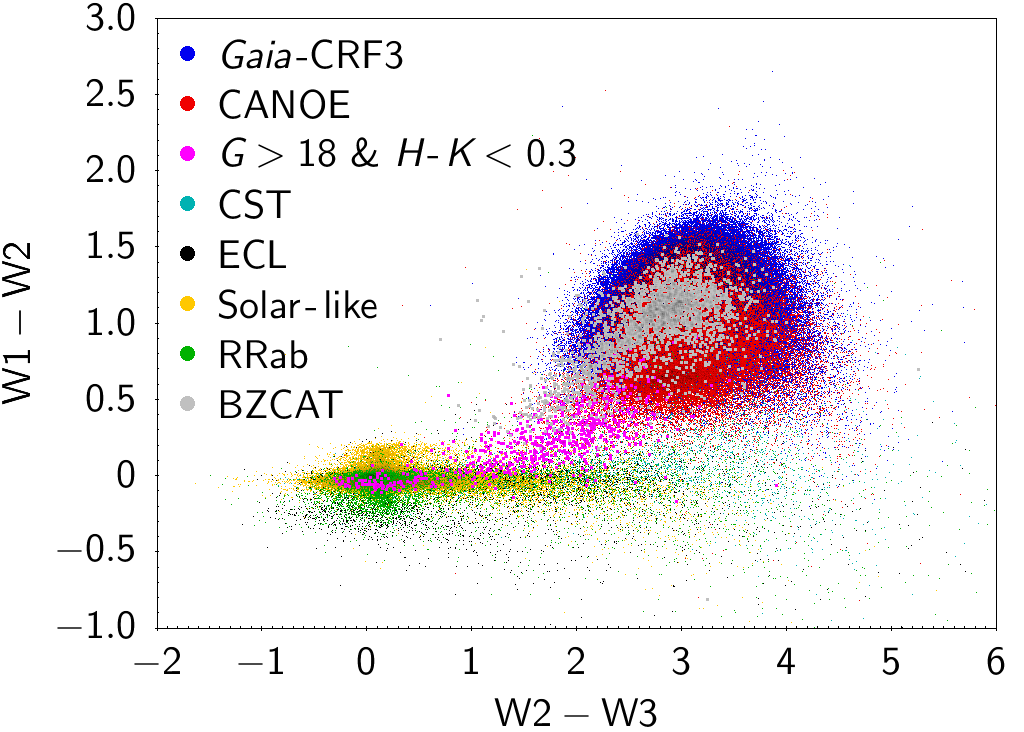}
      \caption{\wise colour-colour diagram. The variable AGN candidates included in the \gaia-CRF3 sample are marked in blue, while the CANOE objects are in red. Blazars are shown with larger (grey) symbols to highlight the `blazar strip', which extends from the quasar locus to the early-type galaxy region. Different types of stellar objects are also shown \citep[see][]{gavras2022}: constant stars (CST), eclipsing binaries (ECL), solar-like stars (with spots and flares), and ab-type RR~Lyrae stars (RRab).  Sources with $G>18$ and $H-K<0.3$ are discussed in the text.
              }
         \label{fig:wise}
   \end{figure}

Figure~\ref{fig:wise} shows the location of the GLEAN sources in the \wise colour-colour diagram $W1-W2$ versus $W2-W3$, which is known to be a powerful tool to classify sources (see Sect.~\ref{sec:intro}). 
The variable AGN candidates lie in the region where quasars and other types of AGN (e.g., blazars) are expected to be, confirming our selection. In particular, the CANOE sources
are distributed in a somewhat smaller zone, suggesting that our selection procedure was very stringent, in line with the high-purity requirement. The `blazar strip' \citep{massaro2012,raiteri2014}, connecting the locus of quasars with that of early-type galaxies and mostly populated by BL~Lac objects, is clearly traced by sources belonging to the BZCAT5 catalogue.
The plot also includes stellar objects of different types, which largely separate from the AGN candidates.

\begin{figure}
   \centering
   \includegraphics[width=\hsize]{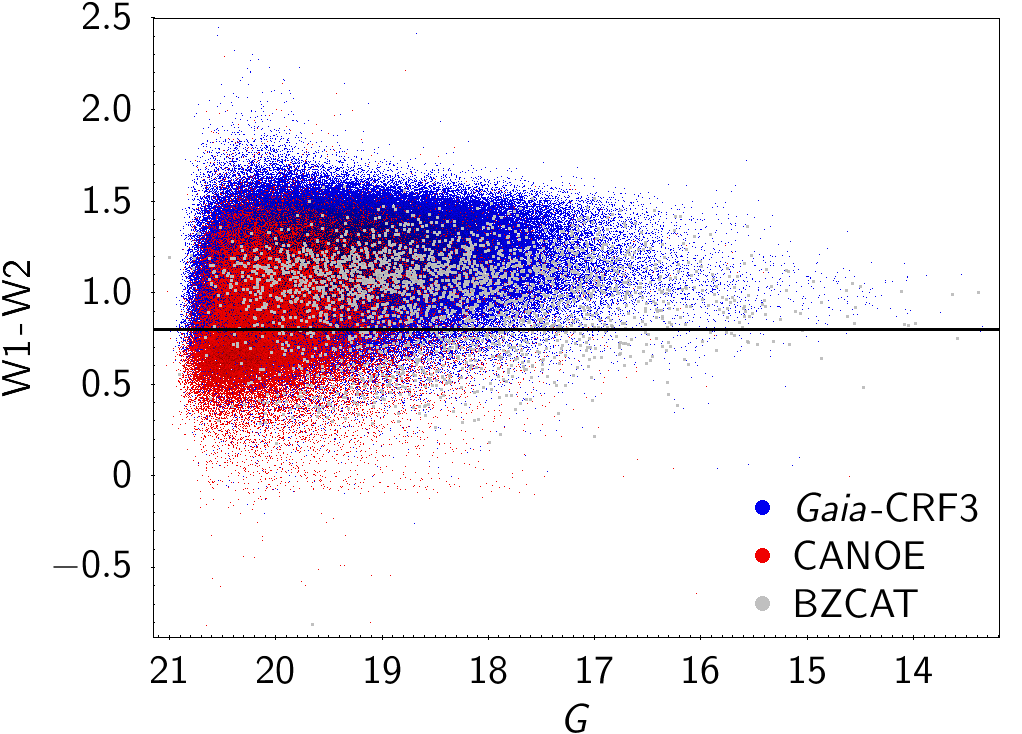}
      \caption{\wise colour index $W1-W2$ versus \gaia \gmag-band magnitude. The horizontal line indicates the threshold $W1-W2=0.8$ above which a source is expected to be a genuine quasar. Most of the sources below this line are very faint objects.
              }
         \label{fig:wise_gaia}
   \end{figure}

There is a fraction of AGN candidates (less than 12\% of GLEAN and 51\% of CANOE sources) with $W1-W2$ less than 0.8, the threshold above which a genuine AGN should lie according to \citet{stern2012}. These sources are mostly faint objects, as shown in Fig.~\ref{fig:wise_gaia}; about 92\% of the GLEAN and 94\% of the CANOE objects with $W1-W2<0.8$ have $G > 19$. Moreover, as noted above, also many blazars, especially BL~Lac objects, have $W1-W2<0.8$. 

\begin{figure}
   \centering
   \includegraphics[width=\hsize]{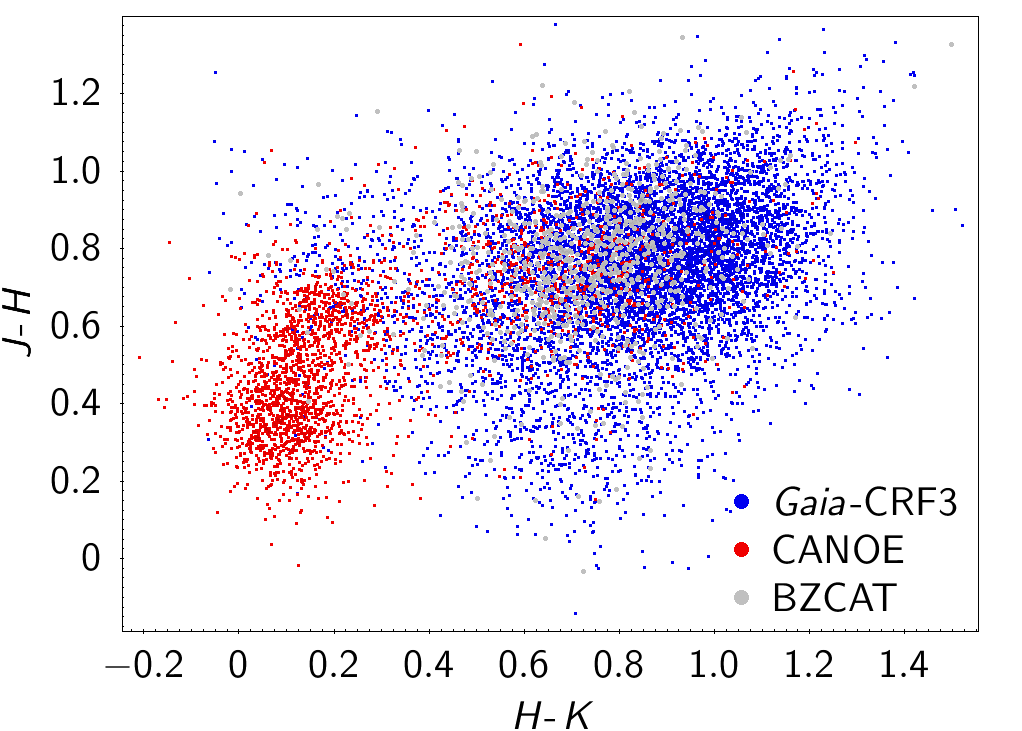}
      \caption{Colour-colour plot of the 11\,215 GLEAN sources with near-IR counterpart in the 2MASS catalogue. 
              }
         \label{fig:2mass}
   \end{figure}

\begin{figure}
   \centering
   \includegraphics[width=\hsize]{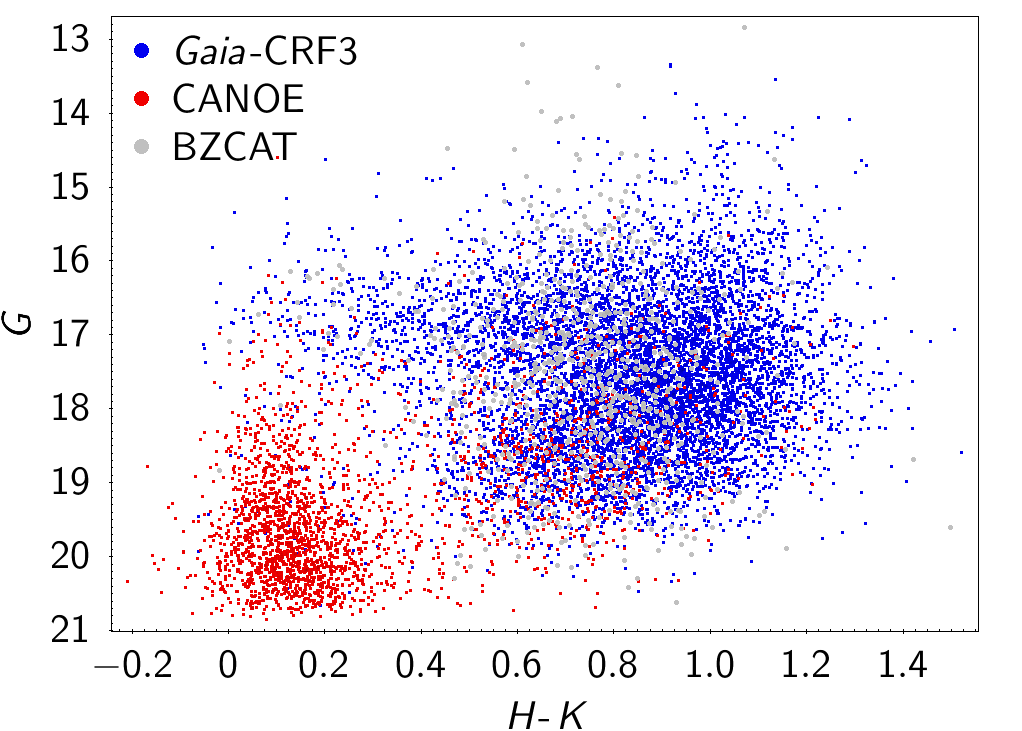}
      \caption{\gaia \gmag magnitude versus the $H-K$ 2MASS colour index.
              }
         \label{fig:2mass_gaia}
   \end{figure}

   Figure~\ref{fig:2mass} shows the colour-colour plot $J-H$ versus $H-K$ of the 11\,215 GLEAN sources (2514 in CANOE) with a near-infrared counterpart in the Two Micron All Sky Survey \citep[2MASS;][]{skrutskie2006} catalogue.
   These counterparts were obtained with a search radius of 3 arcsec and asking for a $\rm SNR>10$.
   We notice a blob of CANOE sources 
   with small values of both $H-K$ and $J-H$. Most of these bluer sources are faint in the \gaia \gmag~band (see Fig.~\ref{fig:2mass_gaia}).
   
   There are 729~CANOE sources with $G>18$ and $H-K<0.3$ which have a \wise counterpart. Most of them lie in a thick strip in the \wise colour-colour diagram (Fig.~\ref{fig:wise}), partly overlapping with the `blazar strip', partly with the region populated by elliptical and spiral galaxies, and partly with stellar sources. This may mean that, notwithstanding all the filters we adopted, our sample is still contaminated by a small fraction of galaxies and stars. Overlaps with galaxies can be expected, as we can have very weak AGN drowned in galaxies.
   A search of the 729~sources in the \gaia~DR3 catalogue of galaxies, containing more than 4.8~million sources (\texttt{galaxy\_candidates} table), yielded 32~matches only. Moreover, 22 out of 729 sources have a radio counterpart, favouring an extragalactic nature.
   
\begin{figure}
   \centering
   \includegraphics[width=8cm]{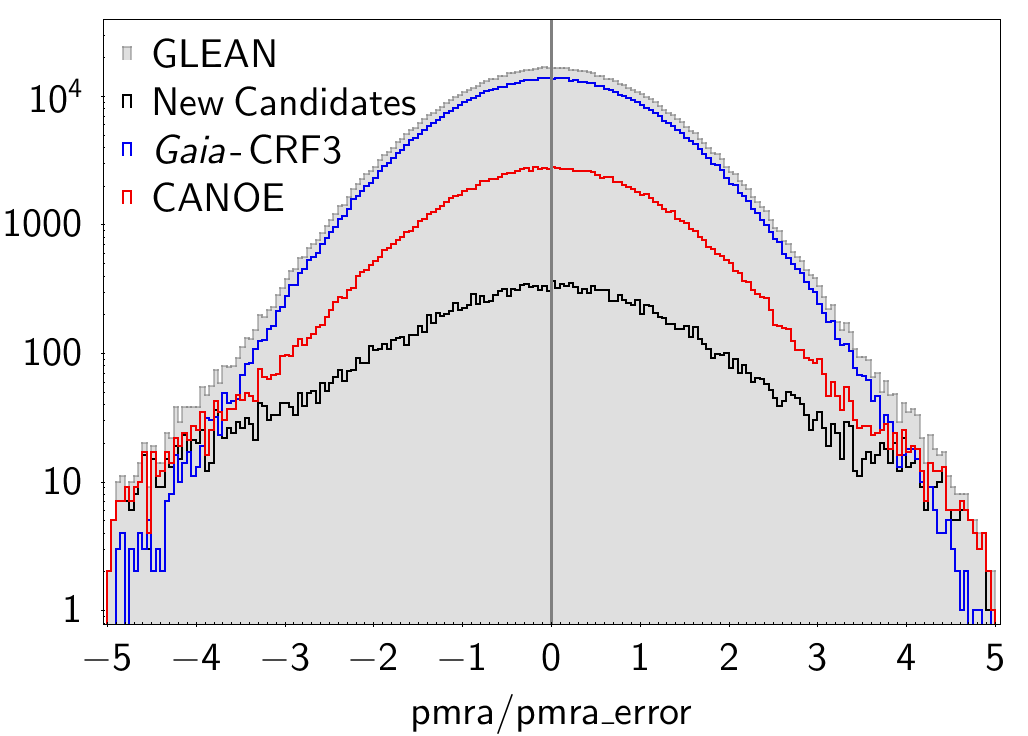}
    \includegraphics[width=8cm]{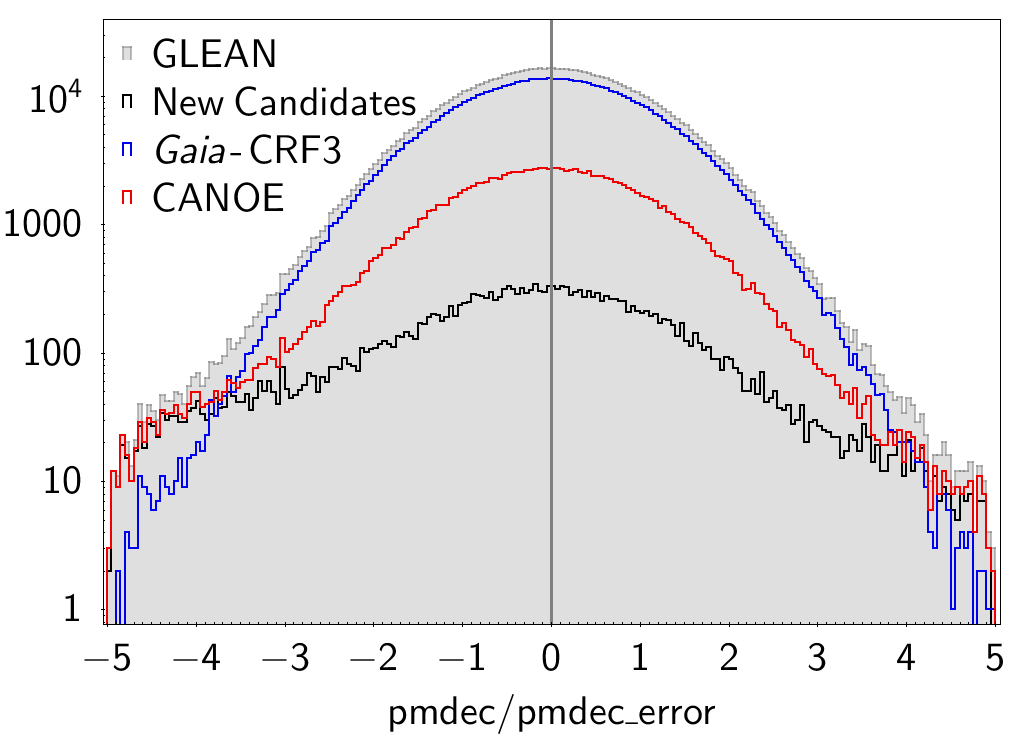}
    \includegraphics[width=8cm]{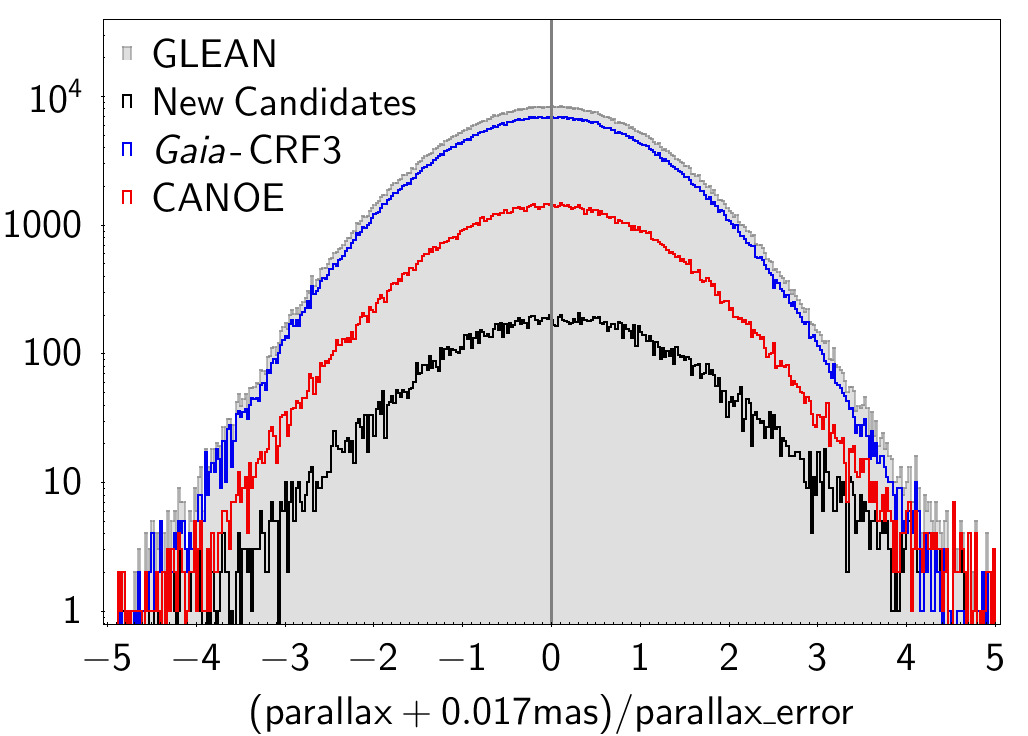}
    \caption{Distributions of proper motion in right ascension (top), proper motion in declination (middle), and parallax (bottom) for the various variable AGN samples discussed in the paper. 
              }
         \label{fig:astro}
   \end{figure}

   The presence of a minor fraction of stellar contaminants is also suggested by the distributions of the \gaia astrometric parameters shown in Fig.~\ref{fig:astro}. A small number excess characterizes the tails (especially the low-side one) of the proper motion distributions of the GLEAN and CANOE samples (and of the newly identified AGN, see Sect.~\ref{sec:copu}) with respect to the \gaia-CRF3 sample \citep[see also][]{klioner2022,liao2021}.

We finally mention that the cross match between the GLEAN sample and the \gaia DR3 \texttt{galaxy\_candidates} table produces 16\,854 overlaps. This is not surprising, as the host galaxy of many nearby AGN is expected to be detectable.

\section{Check for stellar contaminants}
\label{sec:con}

   \begin{figure}
   \centering
   \includegraphics[width=8cm]{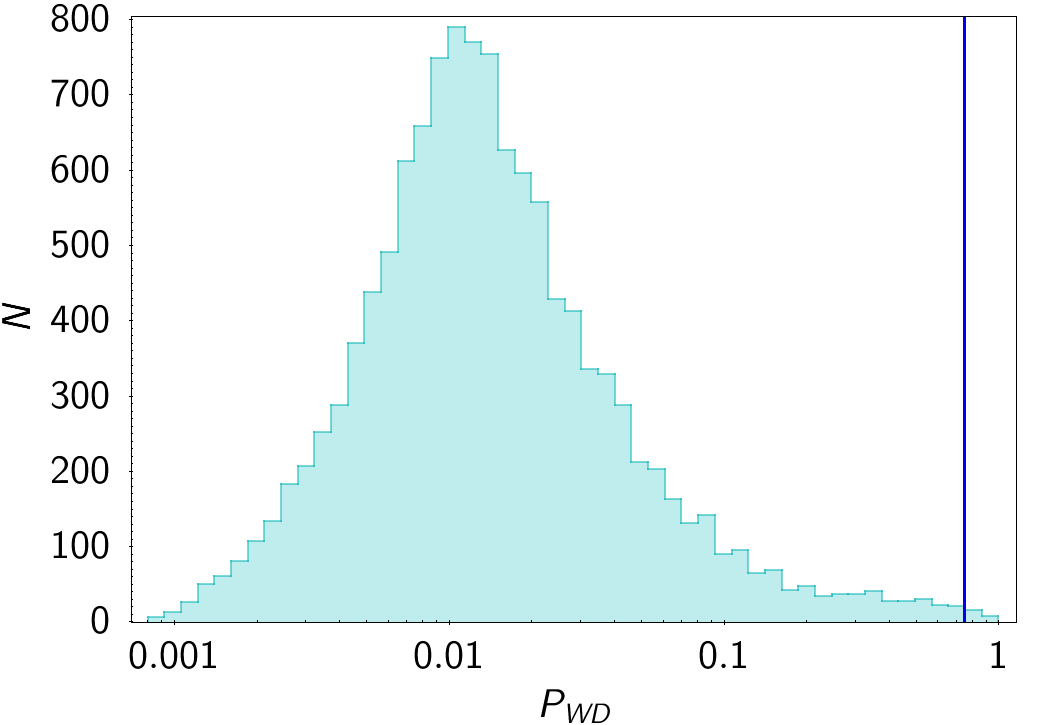}
      \caption{Distribution of $P_{\rm WD}$ for the $\sim$12\,000 sources of the GLEAN sample with a counterpart in the  \gaia~DR2 catalogue of white dwarfs by \citet{gentile2019}. The vertical line highlights $P_{\rm WD}=0.75$, which is the threshold above which a source can be considered a high-confidence WD in the paper.
              }
         \label{fig:wd}
   \end{figure}

   \begin{figure}
   \centering
   \includegraphics[width=8cm]{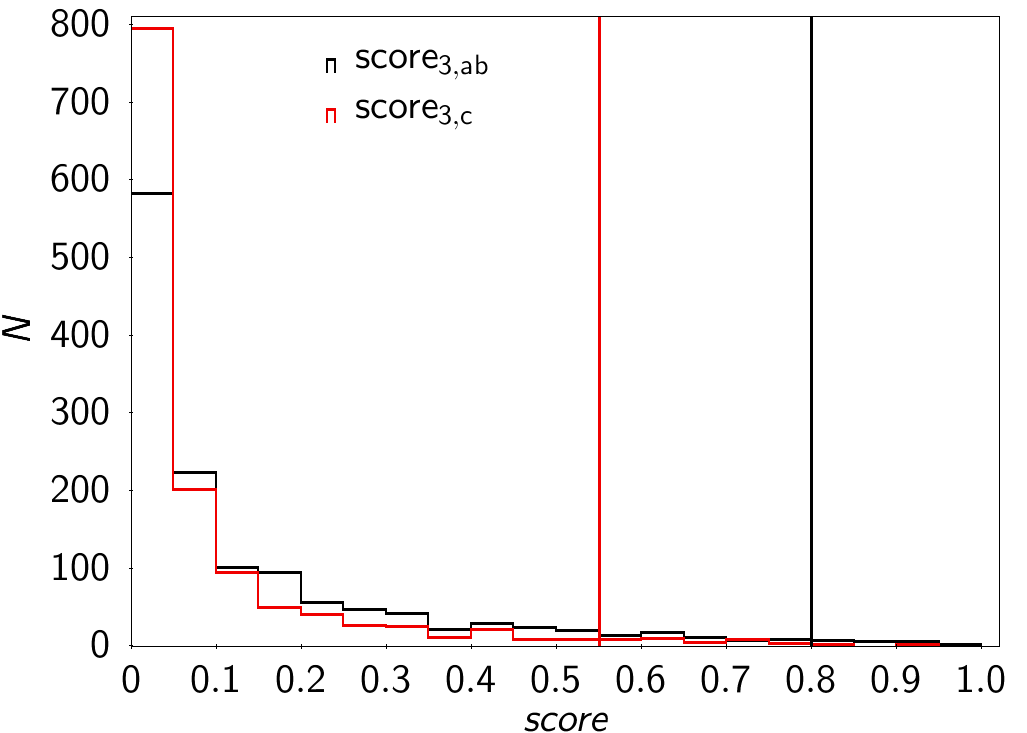}
      \caption{Distribution of the  RRab and RRc classification scores, $\rm score_{3,ab}$ (black line) and $\rm score_{3,c}$ (red line), for the 1319 sources in the GLEAN sample with a counterpart in the catalogue of RR~Lyrae stars by \citet{sesar2017}. Vertical lines indicate the limits  $\rm score_{3,ab}=0.8$ and $\rm score_{3,c}=0.55$, above which sources have a high probability to be RR~Lyrae stars.
              }
         \label{fig:rr}
   \end{figure}

To assess the possible presence of stellar contaminants, we cross-matched the GLEAN sample with various catalogues.

We found that about 12\,156 sources (only 1\,896 in the CANOE sample) are included in the \gaia~DR2 catalogue of white dwarfs (WD) from \citet{gentile2019}. However, a check on the $P_{\rm WD}$ parameter, giving the probability of being a WD, reveals that about 95\% of the sources have $P_{\rm WD}<0.1$ (see Fig.~\ref{fig:wd}), very far from the request $P_{\rm WD}>0.75$ adopted in the paper for high-confidence WD candidates. There are only 24 objects (9 in CANOE) with $P_{\rm WD}>0.75$. 
The inspection of the light curves of the nine CANOE objects with $P_{\rm WD}$ greater than 0.75 reveals long-term variability compatible with an AGN behaviour.

The cross-match with the more recent catalogue of white dwarfs in \gaia~EDR3 by \citet{gentile2021} leads to 55 common objects.
Only 10 of them (1 in CANOE) have $P_{\rm WD} > 0.75$, and their corresponding light curves are compatible with an AGN behaviour.

The cross-match between the GLEAN sources and the PS1 sample of RR~Lyrae stars \citep{sesar2017} yielded 1\,319 common objects (about 388 in CANOE). The distribution of their RRab and RRc classification scores are plotted in Fig.~\ref{fig:rr} and indicates that most sources have a low probability to be RR~Lyrae stars. However, there are 21 objects with $\rm score_{3,ab}>0.8$ and 40 objects with $\rm score_{3,c}>0.55$, which are the limits indicating a high probability to be RR~Lyrae stars. All these 61 objects belong to the \gaia-CRF3 sample and most of them have variability trends in agreement with those of AGN.



We found also 385 sources in GLEAN that are classified as young stellar objects (YSO) in the All-Sky Automated Survey for Supernovae (ASAS-SN) catalogue of variable stars  \citep{jayasinghe2020}, but only 4 with high probability (greater than 0.75) of being YSO.

Further cross-matching with other catalogues of variable stars yielded no significant overlap.

\section{Completeness and Purity}
\label{sec:copu}

We estimate the completeness and purity of the GLEAN sample we have selected, taking into account that this is not a general AGN sample, but a sample of AGN that are observed to be variable. 
The application of the GVD module to \gaia-CRF3 showed that 88\% of AGN are detected as variable in \gaia-DR3.
This is in reasonable agreement with the results of \citet{sesar2007}, who reported that $\ga 90\%$ of the quasars in the Stripe 82 sky region with multiple photometric observations by the SDSS are variable at the 0.03 mag level. 

We first calculated the GLEAN sample completeness
with respect to the SDSS DR16Q v4 catalogue \citep{lyke2020}, which is 99.8\% complete and has only 0.3\%–1.3\% contamination. 
Because the SDSS covers only part of the sky, we selected a wide sampled region, included within $+10 \deg < \rm dec < +50 \deg$ and $130 \deg < \rm ra <220 \deg$. We found 224\,752 DR16Q sources in this area, 145\,669 of which have a \gaia counterpart in the catalogue, and 151\,915 in the \gaia-DR3. In line with the GVD result mentioned above, we assume that 88\% of them are variable, i.e.\ 133\,685 objects.
In the same sky region, we find 62\,696 sources belonging to the GLEAN sample.
Therefore, we can estimate a 47\% completeness of the GLEAN sample when taking the DR16Q catalogue as reference.
Viceversa, there are 38\,650 GLEAN (5\,205 CANOE) sources in the same sky region that are not included in the DR16Q catalogue.

\begin{figure}
   \centering
   \includegraphics[width=8cm]{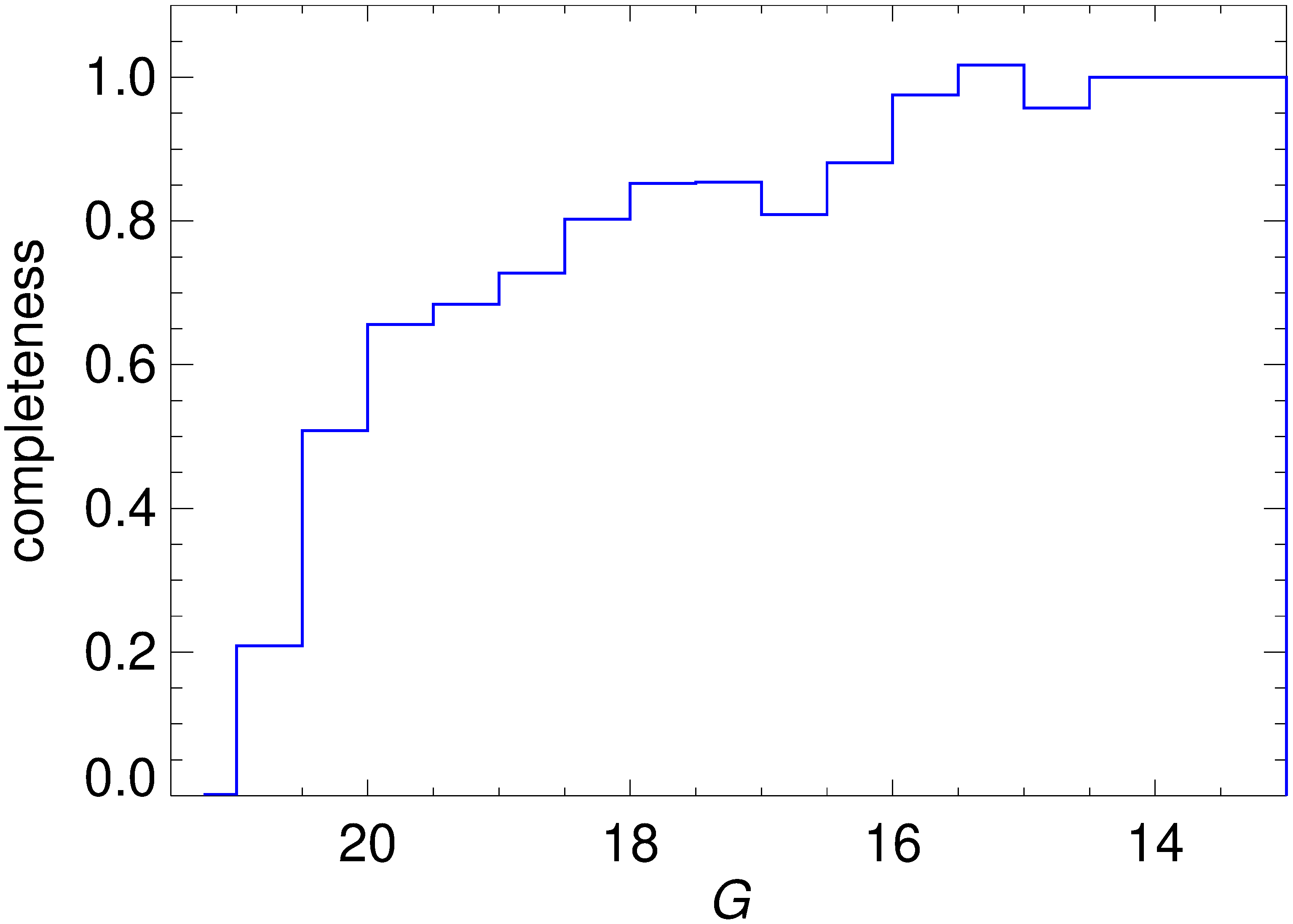}
      \caption{GLEAN sample completeness estimated with respect to the \gaia-CRF3 sample versus \gmag magnitude. 
              }
         \label{fig:cvg}
   \end{figure}

Then we estimate the completeness with respect to the \gaia-CRF3 sample, which we have used as reference for the selection procedure, assuming that it contains genuine AGN.
Actually, the contamination of the \gaia-CRF3 sample is expected to be at most 2\% \citep{klioner2022}.
As before, we assume that the percentage of variable objects is 88\%  of the whole sample, so we can consider that among the 
1\,614\,173 sources in \gaia-CRF3, 1\,420\,472 are variable. On the other side, the number of \gaia-CRF3 source that survived the selection procedure and are present in the GLEAN sample is 722\,211. 
Therefore, we can estimate a completeness of 51\%.
 We analysed the variation of completeness with the \gmag magnitude. Figure~\ref{fig:cvg} shows the ratio between the number of \gaia-CRF3 sources that survived the selection procedure and the number of variable sources in the \gaia-CRF3 sample per magnitude bins. This reveals that the completeness of the final sample is above 90\% for the sources brighter than about \gmag=16 and then decreases in an irregular way with increasing magnitude. It is still about 50\%  at \gmag=20--20.5, and then falls rapidly.

In addition, we estimated the percentage of sources that survived the series of cuts described in Sect.~\ref{sec:sele} both in the case of the \gaia-CRF3 catalogue and for several AGN large external catalogues (with more than 10\,000 sources). The results are reported in 
Table~\ref{tab:cross}.
Columns indicate the catalogue name, the number of sources $N_{\rm cat}$ in each catalogue, the number of matches between the catalogue sources and the initial 34~million variable sources selected by the SOS-AGN module $N_{\rm match,ini}$, the number of matches with the GLEAN sample $N_{\rm match,fin}$, the ratio $N_{\rm match,fin}/N_{\rm match,ini}$, and the number of matches with the CANOE sample $N_{\rm match,new}$.
The ratio $N_{\rm match,fin}/N_{\rm match,ini}$ can be seen as an estimate of the filter survival fraction of the variable sources in that catalogue\footnote{Actually, we did not take into account here that less than 3\,000 sources in GLEAN did not pass the filter selection due to their peculiarities (see Sect.~\ref{sec:further}). However, these represent less than 0.3\% of the sample, so they cannot significantly change the filter survival fraction estimates.}.

The number of \gaia-CRF3 sources in the initial sample of 34~million candidates is 1\,141\,892, of which  722\,211 are included in the GLEAN sample. This gives a filter survival percentage of about 63\%.

 The largest catalogues, containing more than 500\,000 sources, give in general a filter survival percentage roughly between 60\% and 70\%, with an average value of 65\%. This estimate is also in agreement with those inferred by considering the ``QSO" objects in the APOP catalogue and the e-ROSITA AGN catalogue. The filter survival percentage derived from the other smaller catalogues is higher, ranging from about 70\% to almost 80\%.

\begin{table*}
    \caption{Results of the cross-match of the GLEAN and CANOE samples with large AGN catalogues.}\centering
\begin{tabular}{lrrrrrc} 
\hline\hline 
    Catalogue  & $N_{\rm cat}$& $N_{\rm match,ini}$ & $N_{\rm match,fin}$ & Ratio & $N_{\rm match,new}$ & Reference\\
     \hline
    WISE  C75    & 20\,907\,127 & 925\,988 & 602\,782 & 0.65 & 28\,574 &1\\
    WISE R90    &4\,543\,530 & 788\,714 & 526\,388 & 0.67 & 4\,569 &1\\
    \gaia--unWISE    & 2\,734\,464 & 1\,363\,764 & 819\,677 & 0.60 & 108\,345 &2\\
    \gaia-DR2  & 2\,690\,021& 1\,025\,599 &652\,900 & 0.64  & 85\,662 &3\\
    {\bf \gaia-CRF3}  & {\bf 1\,614\,173} & {\bf 1\,141\,892} & {\bf 722\,211} & {\bf 0.63} & {\bf 0} & {\bf 4}\\
    SDSS DR16Q Superset v3    & 1\,440\,615 & 295\,426 & 196\,907 & 0.67 &1\,360 &5\\
    AllWISE AGN    & 1\,354\,775 & 494\,067 &323\,225 & 0.65 & 2\,316 &6\\
    MILLIQUAS  & 1\,115\,619& 411\,742 & 273\,422& 0.66 &5\,545 &7\\
    SDSS DR16Q v4    & 750\,414 & 291\,484 & 195\,946 & 0.67 &1\,201 &5\\
    LQAC5    &592\,809 & 259\,127 & 174\,076 & 0.67  &89 &8\\
    LQRF    & 100\,165 & 81\,560 & 61\,159 & 0.75 & 40 &9\\
    BROS      & 88\,211 & 6\,304 &4\,510 & 0.72 & 304 &10\\
    APOP (QSO) & 86\,821 & 72\,107  & 54\,407 & 0.63 & 33 &11\\
    LAMOST5    & 52\,453& 38\,341 & 29\,188 & 0.76 & 92 &12\\
    2QZ  & 49\,425& 19\,254 &13\,794 & 0.72 & 184 &13 \\
    e-ROSITA  & 21\,952 & 5\,122 & 3\,319 & 0.65 &289 &14\\
    OCARS    &13\,589 & 6\,541 & 5\,099 & 0.78 &  147 &15\\
    Seyfert  & 11\,101 & 7\,802 & 5\,578 & 0.71  & 26 &16\\

    \hline
    \end{tabular}\\
    \footnotesize{Note: (1) \cite{assef2018}; (2) \cite{shu2019}; (3) \cite{bailer2019}; (4) \cite{klioner2022}; (5) \cite{lyke2020}; (6) \cite{secrest2015}; (7) \cite{flesch2021}; (8) \cite{souchay2019}; (9) \cite{andrei2009}; (10) \cite{itoh2020}; (11) \cite{qi2015}; (12)\cite{yao2019}; (13) \cite{croom2004}; (14) \cite{liu2021}; (15) \cite{malkin2018}; (16) \cite{rakshit2017}}\\
    \label{tab:cross}
\end{table*}

The purity of the GLEAN sample is the number of genuine  variable AGN included in it over the total number of GLEAN objects. 
A lower limit to the purity of the GLEAN sample can be obtained from the ratio between the number of \gaia-CRF3 sources in the sample and the total number of sources in the sample, which is around 83\%. 
However, as derived from the cross-match with the catalogues in Table~\ref{tab:cross}, 128\,282 of the $\sim$150\,000 CANOE objects are present in other AGN catalogues. This in principle raises the purity lower limit of the GLEAN sample to about 97\%.
However, since we cannot exclude that the common sources still include contaminants, we conservatively estimate the sample purity to be around 95\%.

\begin{figure}
   \centering
   \includegraphics[width=\hsize]{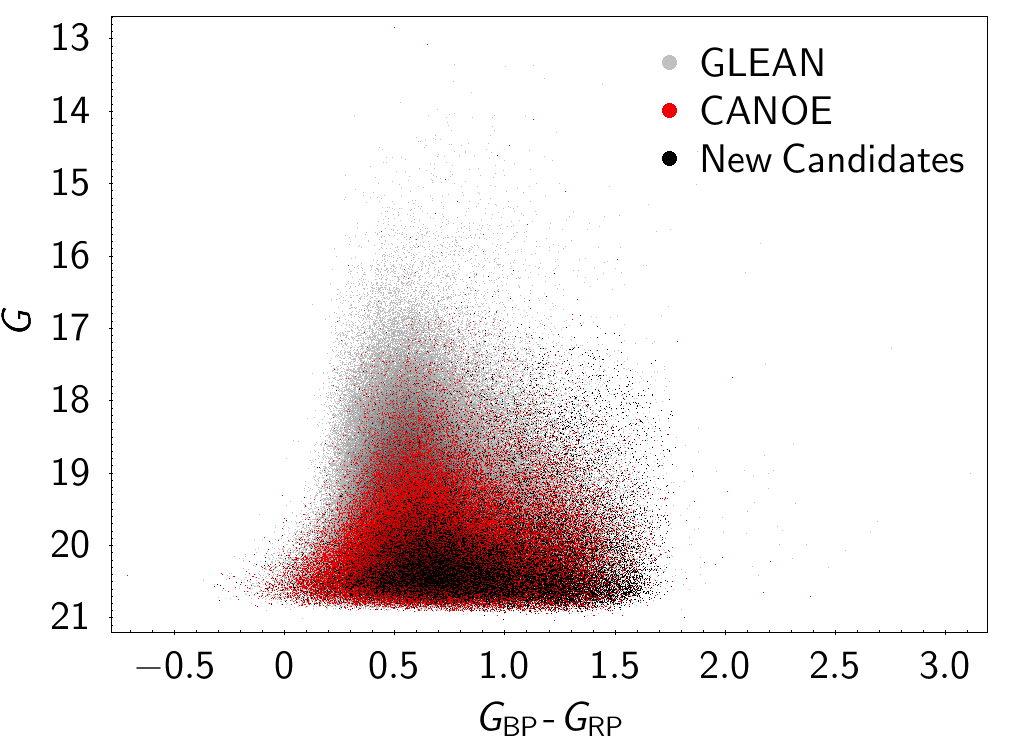}
      \caption{\gaia \gmag versus $G_{\rm BP}-G_{\rm RP}$ colour index for the sources in the GLEAN and CANOE samples, and for the new variable AGN candidates.}
         \label{fig:gcol}
   \end{figure}

From the cross-match with the AGN catalogues in Table \ref{tab:cross} we found that 21\,735 sources are new AGN candidates.
The distribution of astrometric parameters of these new AGN candidates is shown in Fig.~\ref{fig:astro}, while Fig.~\ref{fig:gcol} displays their colour-magnitude diagram, \gmag versus $G_{\rm BP}-G_{\rm RP}$. 
The new sources approximatively cover the same range of $G_{\rm BP}-G_{\rm RP}$ colour indices as the GLEAN and CANOE objects, but they lie among the faintest sources and tend to avoid the region of the bluest colours. The excess in large negative proper motions, previously discussed, appears to be largely due to these new sources, and the effect could be related to their faintness, though we cannot rule out a certain percentage of stellar contamination.

\section{Cross-match with radio catalogues}
\label{sec:radio}

\begin{figure}
   \centering
   \includegraphics[width=\hsize]{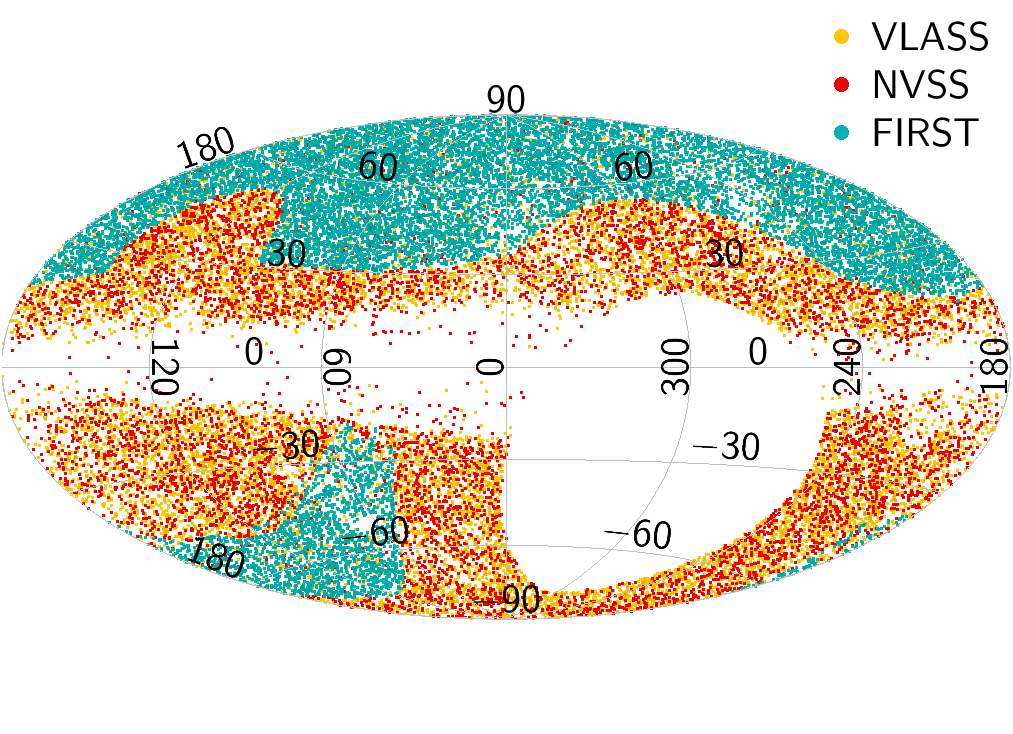}
      \caption{Distribution on the sky of the radio counterparts of the GLEAN sources in the FIRST (cyan), NVSS (red), and VLASS (orange) catalogues, in Galactic coordinates.
              }
         \label{fig:radio_cat}
   \end{figure}

\begin{figure}
   \centering
   \includegraphics[width=\hsize]{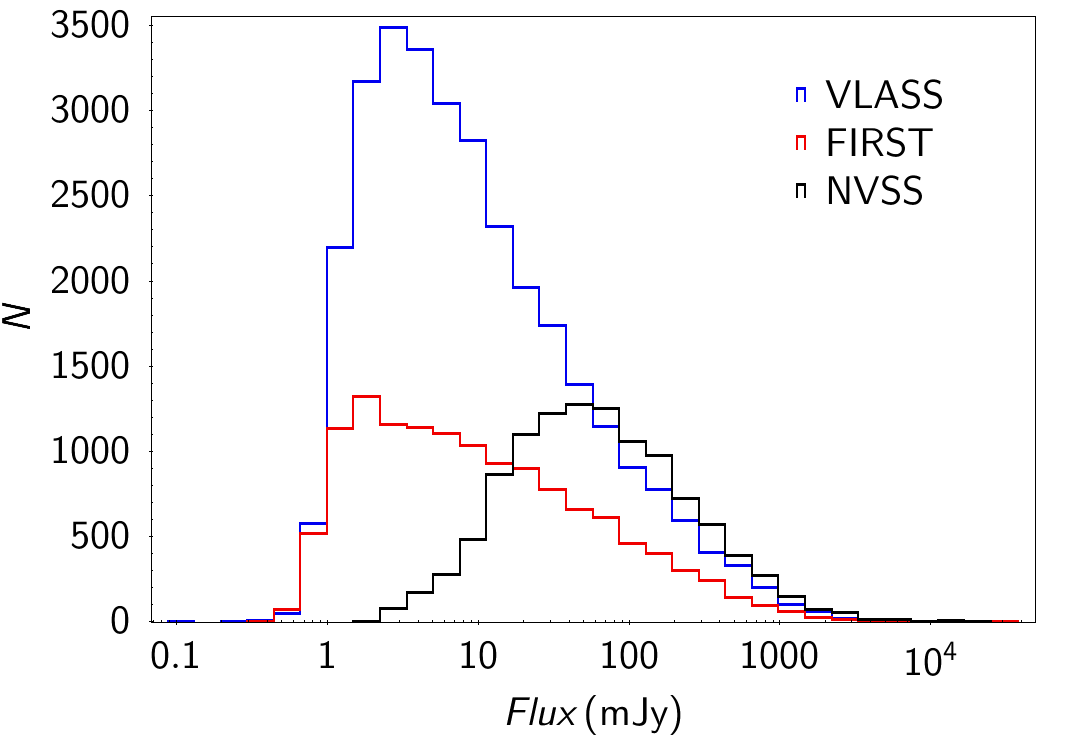}
      \caption{Radio flux densities (mJy) of the counterparts of the GLEAN sources in the FIRST (red), NVSS (black), and VLASS (blue) catalogues.
              }
         \label{fig:radio_flux}
   \end{figure}

As mentioned in Sect.~\ref{sec:intro}, a fraction of AGN are radio-loud. This fraction is generally assumed to be around 10\%, but actually diminishes with increasing redshift and decreasing luminosity \citep{jiang2007,kratzer2015}.
We cross-matched the GLEAN sample with the catalogues of the radio sky surveys FIRST \citep{gordon2021}, NVSS \citep{condon1998}, and VLASS \citep{lacy2020}, using a 1.5~arcsec radius.
Table~\ref{tab:radio} shows, 
for each catalogue, the observing radio frequency, the percentage of the sky covered, the number of objects  $N_{\rm cat}$ in the catalogue, and the number of GLEAN sources $N_{\rm cross}$ with a radio counterpart.
The distribution on the sky of the GLEAN-radio pairs is plotted in Fig.~\ref{fig:radio_cat}.
Fig.~\ref{fig:radio_flux} shows the distribution of radio fluxes, highlighting the greater depth of the FIRST and VLASS catalogues with respect to NVSS and the much larger number of objects in the VLASS.
The number of non-duplicated variable AGN candidates with radio counterparts is 33\,706, which represents about 4\% of the GLEAN sample.

\begin{table}
\caption{Results of the cross-match between the GLEAN sources and radio catalogues}\label{tab:probability}
\label{tab:radio}
\centering
\begin{tabular}{lccrr} 
\hline\hline             
Name & Band (GHz) & Sky  (\%) & $N_{\rm cat}$ & $N_{\rm cross}$ \\ 
\hline
FIRST & 1.4 & 25.6 & 946\,432 & 13\,133 \\ 
VLASS & 3.0 & 82 & 3\,381\,277 & 31\,378  \\ 
NVSS & 1.4 & 82 & 1\,773\,484 & 11\,041 \\ 
\hline
\end{tabular}
\end{table}

      \begin{figure}
   \centering
   \includegraphics[width=\hsize]{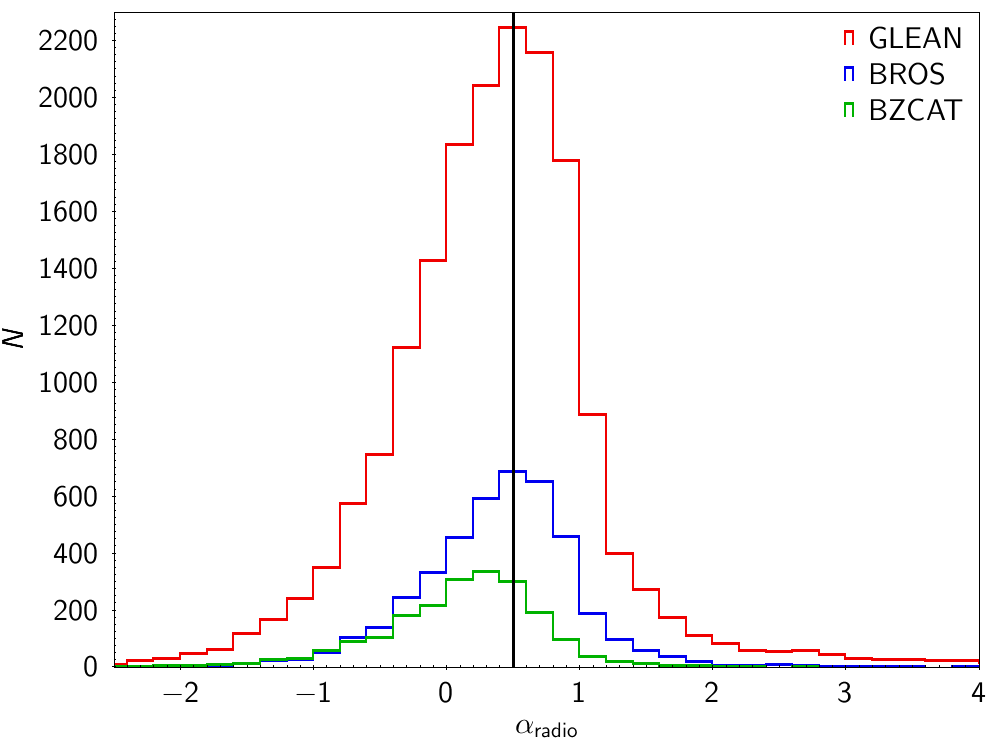}
      \caption{Distribution of the 1.4--3.0~GHz spectral index for the 17\,399 radio counterparts of the GLEAN variable AGN (red), for the 2058 of them that are included in the BZCAT5 catalogue of confirmed blazars (green), and for the 4209 blazar candidates in the BROS catalogue (blue). The vertical line indicates the value 0.5 below which a radio spectrum is defined as `flat'.
              }
         \label{fig:index_radio}
   \end{figure}

Under the assumption that the 1.4--3.0~GHz spectrum can be approximated by a power-law $F_\nu \propto \nu^{-\alpha}$, we calculated the 1.4--3.0~GHz spectral index for the 17\,399 counterparts of the GLEAN sources with radio data in both bands. When 1.4 GHz information from both FIRST and NVSS was available, we chose the latter, so that we used NVSS for about 58\% of the sources. 
The spectral index is plotted in Fig.~\ref{fig:index_radio}; its median value is $0.39$ and the standard deviation 0.88.
The median value does not change significantly if we set a lower limit of 3 mJy or even 10 mJy to the VLASS flux.
By comparing these results to those by \citet{gordon2021}, we found a good agreement, taking into account that we are mostly dealing with compact sources.

If we apply the classical condition $\alpha < 0.5$ to define a flat spectrum \citep[e.g.][]{urry1995}, we find 9949 sources (57\%) with a flat spectrum, which is a distinctive feature of a blazar source. 
Actually, a reliable spectral index for a variable source should be calculated with contemporaneous data in the two bands. Here it is not possible, and this must be kept in mind when evaluating the results. For instance, blazars have flat radio spectra, and indeed $\sim 75\%$ of the 2058 confirmed blazars in the BZCAT5 catalogue for which we could estimate the radio spectral index show values smaller than 0.5 (see Fig.~\ref{fig:index_radio}), but still $\sim 25\%$ of blazars display a steep spectrum. 

Spectral indices with non-contemporaneous data have been used to identify blazar candidates, as in the cases of the CRATES \citep{healey2007} and BROS \citep{itoh2020} catalogues. 
In particular, the selection criterion for the BROS blazars was to have $\alpha_{\rm radio} < 0.6$, as derived from the Fermi 4LAT sources \citep{abdollahi2020}, and the spectral index was obtained by using radio data from 0.15~GHz TGSS \citep{intema2017} and 1.4~GHz NVSS catalogues.
However, among the 4209 BROS expected `flat-spectrum' sources in Fig.~\ref{fig:index_radio}, only 2327 (55\%) have actually a flat spectrum according to our criterion. In the above discussion, we have assumed that the broad-band radio spectrum can be approximated by a power law. Deviations from a power-law SED would modify the above numbers.

We investigated the percentage of radio-loud sources in our sample. The classical definition of a radio-loud source is that $R=F_{\rm 5 GHz}/F_B >10$ \citep[e.g.][]{urry1995}, where $F_{\rm 5 GHz}$ and $F_B$ are the flux densities at 5~GHz and in the optical $B$ band, respectively.
For the 17\,399 sources for which $\alpha_{\rm radio}$ could be estimated, the $F_{\rm 5 GHz}$ flux density was derived from that at 3.0~GHz in the hypothesis that the estimated $\alpha_{\rm radio}$ is fairly describing the 3--5~GHz spectrum too. 

\begin{figure}
   \centering
   \includegraphics[width=\hsize]{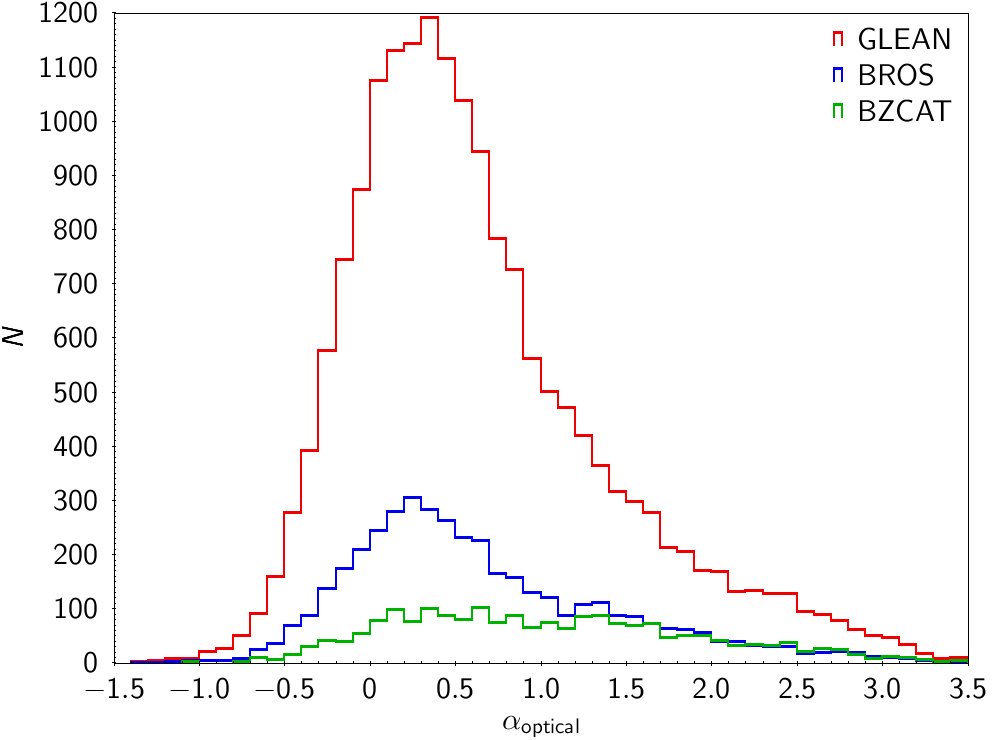}
      \caption{Same as Fig.~\ref{fig:index_radio} for the optical spectral index.
              }
         \label{fig:index_optical}
   \end{figure}

In order to calculate $F_B$, we made the assumption that we can approximate the spectrum with a power law also in the optical. Therefore, we first obtained Johnson-Cousins $V$ and $R$ magnitudes from \gaia\ magnitudes according to the relationships provided by \citet{riello2021}. Then we calculated the corresponding flux densities using the zeropoints by \citet{bessell1998}, and then corrected them for Galactic reddening according to \citet{schlegel1998} and \citet{fitzpatrick1999}. The resulting optical spectral index $\alpha_{\rm opt}$ is shown in Fig.~\ref{fig:index_optical}. The average value is $0.64 \pm 0.77$.
Finally, for each source we derived $F_B$ from $F_V$ using its own $\alpha_{\rm opt}$.

\begin{figure}
   \centering
   \includegraphics[width=\hsize]{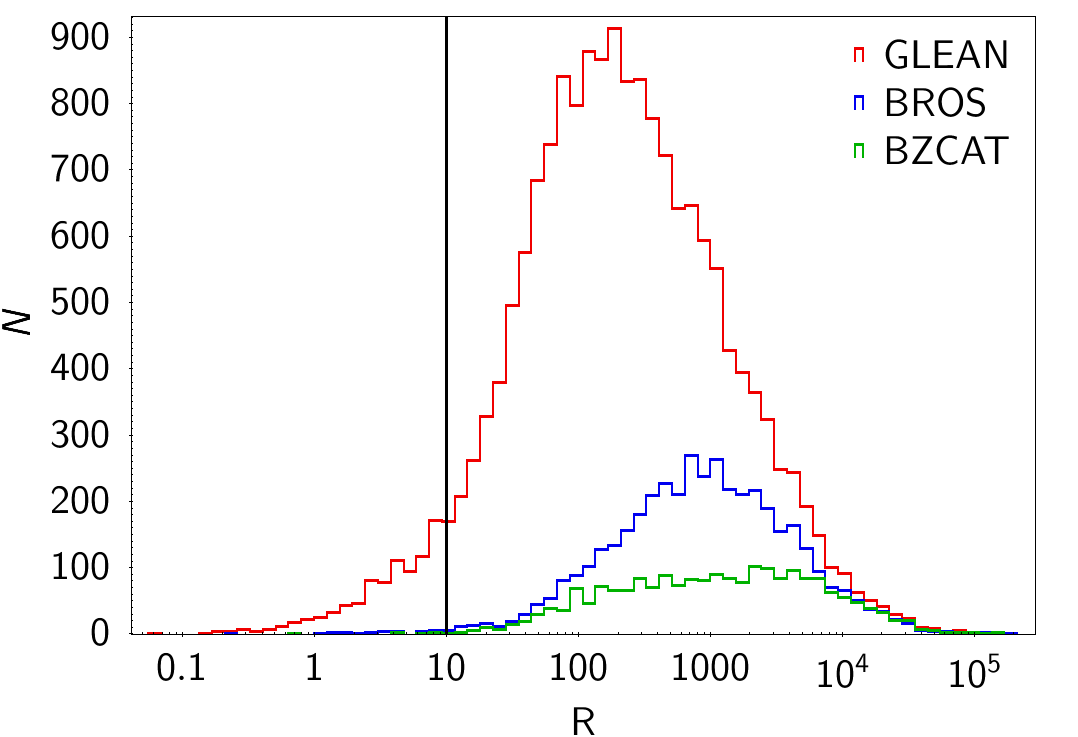}
      \caption{As Fig.~\ref{fig:index_radio} for the distribution of
the radio-loudness parameter $R$. The vertical line indicates the value $R=10$, above which a source is classically defined as `radio-loud'.
              }
         \label{fig:radio_loud}
   \end{figure}

The distribution of the radio-loudness parameter $R$ is shown in Fig.~\ref{fig:radio_loud}. The number of radio-loud sources is 16\,459, which represents 95\% of the 17\,399 sources for which we could calculate a radio spectral index.

If we simply generalized this result to all sources with a radio counterpart, taking into account the different sky coverage of \gaia\ with respect to the radio surveys, we would infer that the number of radio-loud sources in our GLEAN sample is of the order of 4\%.

\section{Lensed quasars}
\label{sec:lens}

The GLEAN sample includes more than a hundred known gravitationally lensed quasars\footnote{For a complete list, see the Gravitationally Lensed Quasar Database at \url{https://research.ast.cam.ac.uk/lensedquasars/}}.

\begin{figure*}
   \centering
   \includegraphics[width=4.5cm]{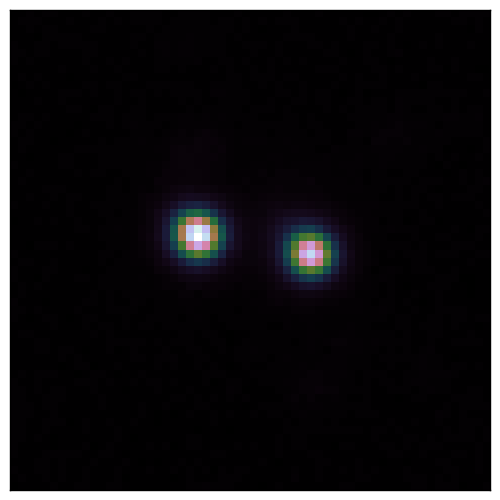}
   \includegraphics[width=6.5cm]{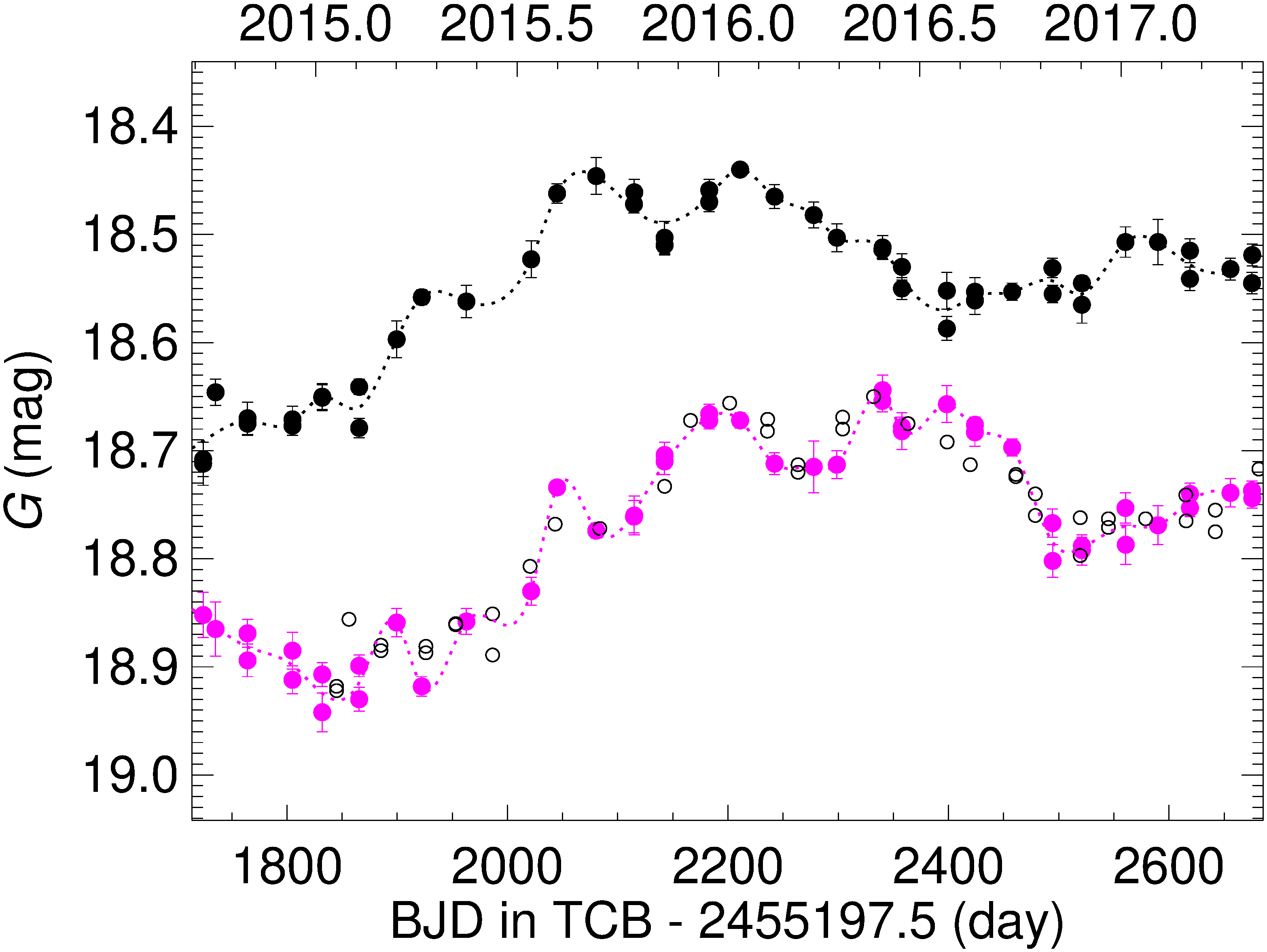}
   \includegraphics[width=6.5cm]{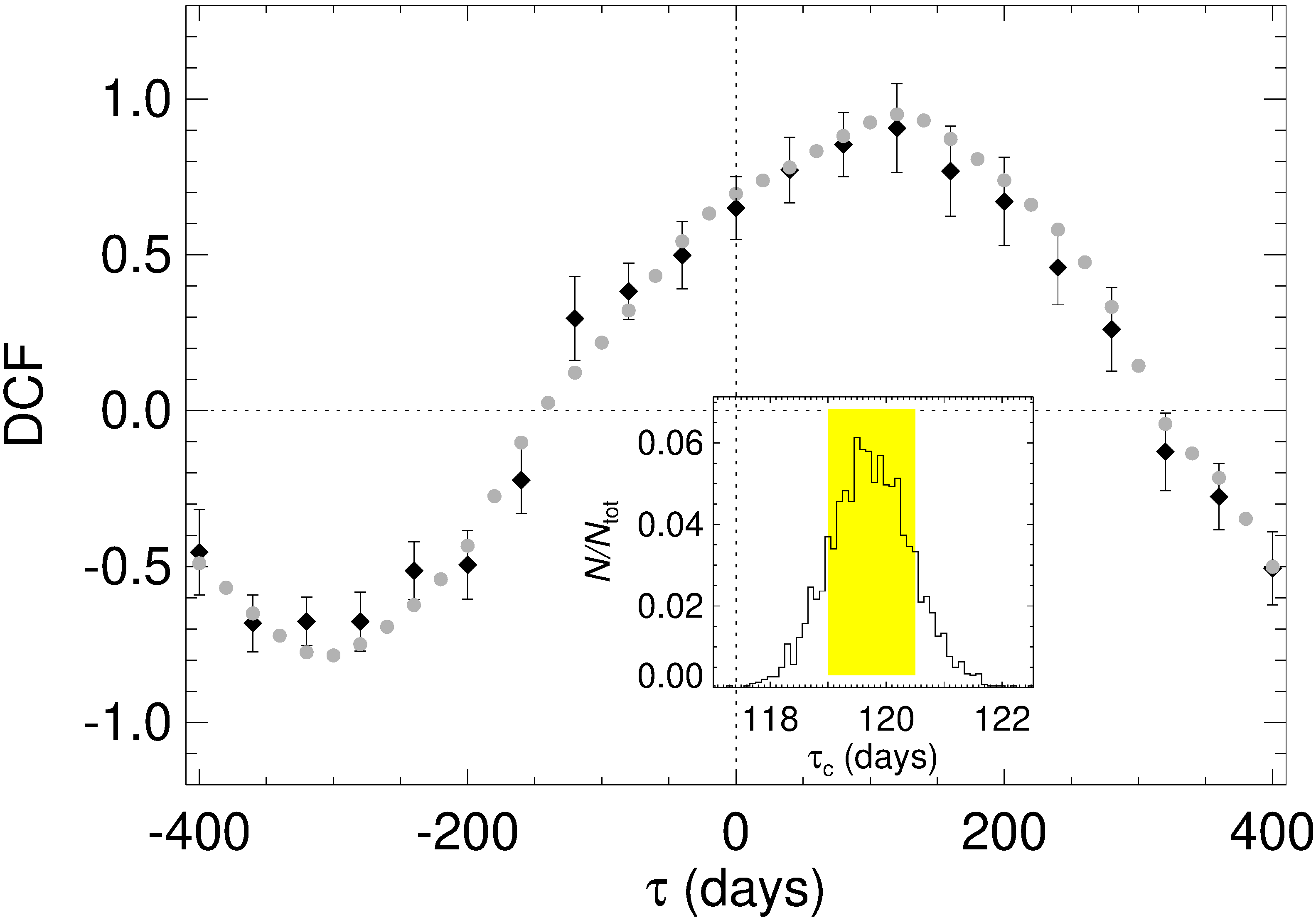}
         \caption{Left: Dark Energy Survey (DES) $g$-band image of the lens system DESJ0501-4118. Middle: \gaia \gmag-band light curves of image~1 (black dots) and image~2 (magenta dots); the empty circles represent the image~1 light curve shifted in time by 121~days and in brightness by 0.21~mag to match the behaviour of image~2. Dotted lines are cubic spline interpolations through the 30-d binned light curves. Right: DCF between the image~1 and image~2 light curves (black diamonds) and between their splines (grey dots); they indicate a time delay of $\sim$120~days of the flux variations of image~2 with respect to those of image~1. The inset shows the result of 3000 Monte Carlo DCF simulations; the yellow strip highlights the interval of the time lag centroid values including 68\% of cases ($1 \sigma$).          }
         \label{fig:lens}
\end{figure*}

We investigated the possibility of deriving robust measurements of the time lag between the observed flux variations corresponding to the various images of a lensed quasar, which is the first step that can lead to the determination of the value of the Hubble constant \citep[e.g.,][and references therein]{tewes2013,wong2020}. This is a difficult task, because quasars are characterized by smooth variability on time-scales of months and because microlensing by the stars of the lensing galaxy can produce additional features, which are different in the light curves of the various images.
Long-term monitoring with good sampling is thus necessary to match the light curve of one image with that of another image through the application of the right shift in time and brightness. The detection of well-defined characteristic patterns of variability substantially improves the time lag estimate. 
We found such an example in the double-lensed quasar DESJ0501-4118 \citep{lemon2019}, which is shown in Fig.~\ref{fig:lens}.
The characteristic variability behaviour, with a double bump in the light curve of the brighter image (image~1), which can be recognized in the light curve of the fainter image (image~2) after some delay, makes the possibility of a robust time lag determination promising.
Microlensing effects by stars within the lensing galaxy seem important here and, as mentioned before, can explain differences between the two light curves that cannot be accounted for by shifts in time and magnitude.
 Because of these effects, the simple application of a  discrete correlation function \citep[DCF;][]{edelson1988,hufnagel1992}, a method which was specifically designed to cross-correlate unevenly-sampled data trains, gives somewhat unstable results, which depend on the DCF time lag bin.
A detailed treatment of the microlensing effects is beyond the scope of this paper. 
However, an estimate of the time lag can be obtained by considering cubic spline interpolations through the binned light curves, which highlight the long-term trend while smoothing the short-term oscillations.

We first calculated the cubic spline interpolation through the 30-day binned light curve of image~1 ($G_{1,\rm spline}$).
Then we shifted this spline by a quantity $\tau$ in time and a quantity $\zeta$ in magnitude to find the values of these two parameters that lead to the best match with the light curve of image~2  - whose data points $G_2(t_i)$ have errors $\sigma_2(t_i)$ - i.e., that minimize the reduced chi-squared (with $\nu$ degrees of freedom):

$$\frac{\chi^2}{\nu}=\frac{1}{N-1}\sum_{ij} \left(\frac{G_{1,\rm spline}(t_j+\tau)+\zeta-G_2(t_i)}{\sigma_2(t_i)}\right)^2$$
for all $N$ pairs $ij$ of points that are separated by no more than 5~days, i.e.\ for which $|t_j+\tau-t_i|<5$.
The result was $\tau=121 \, \rm d$ and $\zeta=0.21 \, \rm mag$.
Figure~\ref{fig:lens} shows the match between the data of image~1 and image~2 when the former is shifted by 121~days and 0.21~mag. Decreasing the spline bin to 20~d does not change the results, while increasing it to 40~d leads to $\tau=120 \, \rm d$, but in both cases the $\chi^2/\nu$ increases with respect to the 30-d bin. 

The DCF between the two light curves and the one between the two splines (see Fig.~\ref{fig:lens}) show a peak at $\tau=120 \, \rm d$, but the centroid indicates a somewhat smaller time lag: about 117~d for the DCF on the light curves, and 119~d for that on the splines. 

To determine the uncertainty on the time lag, we ran 3000 `flux randomization/random subset selection' Monte Carlo simulations \citep{peterson1998,raiteri2003}. The distribution of the lag centroids is shown in Fig.~\ref{fig:lens}; in 68\% of cases ($1 \, \sigma$) the delay is between 119 and 120.4~days.
Altogether, we conclude that the brightness variations of image~2 follow those of image~1 with a time lag of 119--121~d.

\section{Summary and conclusions}
\label{sec:fine}
We have presented the \gaia SOS-AGN module included in the variability analysis pipeline, and the subsequent procedure to select variable AGN candidates.
The result is a high-purity variable AGN sample (GLEAN), including more than 872\,000 sources. 
Starting from initial requirements (more than 20~FoV transits in the \gmag band light curve and having some variability metrics defined), the following filters were tailored on the \gaia-CRF3 sample and included cuts on the structure function index, the \citet{butler2011} statistics, colour indices, parallax, proper motion, and environment density.
We also introduced filters on the effect of scan angle variations and on the GVD variability probability, to avoid contamination by artificially variable nearby galaxies. We notice that the upstream module of General Supervised Classification includes as target categories galaxies, detectable from their spurious variable signal, stars (in 23 types) and AGN. Sources with spurious variability due to scan-angle variations are expected to be assigned to the galaxy class, with the AGN class mostly retrieving the extragalactic sources dominated by the emission from an active nucleus \citep{eyer2022,rimoldini2022}. Moreover, the selection of variable AGN presented in this paper is tailored on the \gaia-CRF3 sample, which includes mostly AGN dominated by the nucleus, rather than by the galaxy.
In addition, our sources are characterized by good astrometric solutions, while galaxies dominated by artificial variability have in general astrometric solutions of lower quality. We also note that
a small contribution from the artificial variability of the host galaxy would in any case be diluted from the AGN contribution. In conclusion, we expect that only a minor fraction of sources in our sample may be affected in a sensitive way by artificial variability introduced by an extended host galaxy.
All filters are based on \gaia data only.
The GLEAN sample has a 47\% completeness when we take the SDSS DR16Q quasar catalogue as a reference, assuming that 88\% of the sources are variable.
The completeness estimated as the percentage of \gaia-CRF3 variable AGN identified by our selection procedure with respect to those in the complete sample is 51\%. We found that this value strongly depends on magnitude. 
We further evaluated the specific impact of the series of filters applied to
the sources selected by the SOS-AGN module. When considering the \gaia-CRF3 sample, the filter survival percentage is about 63\%, i.e.\ the cuts are responsible for the removal of 37\% of the candidates. Taking into account other large AGN catalogues, the cut survival percentage ranges between about 60\% and 80\%.
The purity of the GLEAN sample is conservatively estimated to be higher than 95\%.
This result comes from both the comparison with other AGN catalogues and a careful investigation of possible contaminants.

We have discussed the properties of the selected AGN, complementing \gaia data with data from near-IR, mid-IR, and radio surveys. In particular, we have estimated that about 4\% of the selected sources are radio-loud according to the classical definition.

Finally, we have shown the potentiality of \gaia\ light curves to estimate the time lags between the flux variations of the multiple images of lensed quasars. This goal would be more easily achieved by merging \gaia data with other datasets.

\section*{Acknowledgements\label{sec:acknowl}}
\addcontentsline{toc}{chapter}{Acknowledgements}
This work presents results from the European Space Agency (ESA) space mission \gaia. \gaia\ data are being processed by the \gaia\ Data Processing and Analysis Consortium (DPAC). Funding for the DPAC is provided by national institutions, in particular the institutions participating in the \gaia\ MultiLateral Agreement (MLA). The \gaia\ mission website is \url{https://www.cosmos.esa.int/gaia}. The \gaia\ archive website is \url{https://archives.esac.esa.int/gaia}.
Acknowledgements are given in Appendix~\ref{ssec:appendixA}


\bibliographystyle{aa} 
\bibliography{agn} 

\begin{appendix}
\section{}\label{ssec:appendixA}
This work presents results from the European Space Agency (ESA) space mission \gaia. \gaia\ data are being processed by the \gaia\ Data Processing and Analysis Consortium (DPAC). Funding for the DPAC is provided by national institutions, in particular the institutions participating in the \gaia\ MultiLateral Agreement (MLA). The \gaia\ mission website is \url{https://www.cosmos.esa.int/gaia}. The \gaia\ archive website is \url{https://archives.esac.esa.int/gaia}.

The \gaia\ mission and data processing have financially been supported by, in alphabetical order by country:
\begin{itemize}
\item the Algerian Centre de Recherche en Astronomie, Astrophysique et G\'{e}ophysique of Bouzareah Observatory;
\item the Austrian Fonds zur F\"{o}rderung der wissenschaftlichen Forschung (FWF) Hertha Firnberg Programme through grants T359, P20046, and P23737;
\item the BELgian federal Science Policy Office (BELSPO) through various PROgramme de D\'{e}veloppement d'Exp\'{e}riences scientifiques (PRODEX) grants and the Polish Academy of Sciences - Fonds Wetenschappelijk Onderzoek through grant VS.091.16N, and the Fonds de la Recherche Scientifique (FNRS), and the Research Council of Katholieke Universiteit (KU) Leuven through grant C16/18/005 (Pushing AsteRoseismology to the next level with TESS, GaiA, and the Sloan DIgital Sky SurvEy -- PARADISE);  
\item the Brazil-France exchange programmes Funda\c{c}\~{a}o de Amparo \`{a} Pesquisa do Estado de S\~{a}o Paulo (FAPESP) and Coordena\c{c}\~{a}o de Aperfeicoamento de Pessoal de N\'{\i}vel Superior (CAPES) - Comit\'{e} Fran\c{c}ais d'Evaluation de la Coop\'{e}ration Universitaire et Scientifique avec le Br\'{e}sil (COFECUB);
\item the Chilean Agencia Nacional de Investigaci\'{o}n y Desarrollo (ANID) through Fondo Nacional de Desarrollo Cient\'{\i}fico y Tecnol\'{o}gico (FONDECYT) Regular Project 1210992 (L.~Chemin);
\item the National Natural Science Foundation of China (NSFC) through grants 11573054, 11703065, and 12173069, the China Scholarship Council through grant 201806040200, and the Natural Science Foundation of Shanghai through grant 21ZR1474100;  
\item the Tenure Track Pilot Programme of the Croatian Science Foundation and the \'{E}cole Polytechnique F\'{e}d\'{e}rale de Lausanne and the project TTP-2018-07-1171 `Mining the Variable Sky', with the funds of the Croatian-Swiss Research Programme;
\item the Czech-Republic Ministry of Education, Youth, and Sports through grant LG 15010 and INTER-EXCELLENCE grant LTAUSA18093, and the Czech Space Office through ESA PECS contract 98058;
\item the Danish Ministry of Science;
\item the Estonian Ministry of Education and Research through grant IUT40-1;
\item the European Commission’s Sixth Framework Programme through the European Leadership in Space Astrometry (\href{https://www.cosmos.esa.int/web/gaia/elsa-rtn-programme}{ELSA}) Marie Curie Research Training Network (MRTN-CT-2006-033481), through Marie Curie project PIOF-GA-2009-255267 (Space AsteroSeismology \& RR Lyrae stars, SAS-RRL), and through a Marie Curie Transfer-of-Knowledge (ToK) fellowship (MTKD-CT-2004-014188); the European Commission's Seventh Framework Programme through grant FP7-606740 (FP7-SPACE-2013-1) for the \gaia\ European Network for Improved data User Services (\href{https://gaia.ub.edu/twiki/do/view/GENIUS/}{GENIUS}) and through grant 264895 for the \gaia\ Research for European Astronomy Training (\href{https://www.cosmos.esa.int/web/gaia/great-programme}{GREAT-ITN}) network;
\item the European Cooperation in Science and Technology (COST) through COST Action CA18104 `Revealing the Milky Way with \gaia (MW-Gaia)';
\item the European Research Council (ERC) through grants 320360, 647208, and 834148 and through the European Union’s Horizon 2020 research and innovation and excellent science programmes through Marie Sk{\l}odowska-Curie grant 745617 (Our Galaxy at full HD -- Gal-HD) and 895174 (The build-up and fate of self-gravitating systems in the Universe) as well as grants 687378 (Small Bodies: Near and Far), 682115 (Using the Magellanic Clouds to Understand the Interaction of Galaxies), 695099 (A sub-percent distance scale from binaries and Cepheids -- CepBin), 716155 (Structured ACCREtion Disks -- SACCRED), 951549 (Sub-percent calibration of the extragalactic distance scale in the era of big surveys -- UniverScale), and 101004214 (Innovative Scientific Data Exploration and Exploitation Applications for Space Sciences -- EXPLORE);
\item the European Science Foundation (ESF), in the framework of the \gaia\ Research for European Astronomy Training Research Network Programme (\href{https://www.cosmos.esa.int/web/gaia/great-programme}{GREAT-ESF});
\item the European Space Agency (ESA) in the framework of the \gaia\ project, through the Plan for European Cooperating States (PECS) programme through contracts C98090 and 4000106398/12/NL/KML for Hungary, through contract 4000115263/15/NL/IB for Germany, and through PROgramme de D\'{e}veloppement d'Exp\'{e}riences scientifiques (PRODEX) grant 4000127986 for Slovenia;  
\item the Academy of Finland through grants 299543, 307157, 325805, 328654, 336546, and 345115 and the Magnus Ehrnrooth Foundation;
\item the French Centre National d’\'{E}tudes Spatiales (CNES), the Agence Nationale de la Recherche (ANR) through grant ANR-10-IDEX-0001-02 for the `Investissements d'avenir' programme, through grant ANR-15-CE31-0007 for project `Modelling the Milky Way in the \gaia era’ (MOD4Gaia), through grant ANR-14-CE33-0014-01 for project `The Milky Way disc formation in the \gaia era’ (ARCHEOGAL), through grant ANR-15-CE31-0012-01 for project `Unlocking the potential of Cepheids as primary distance calibrators’ (UnlockCepheids), through grant ANR-19-CE31-0017 for project `Secular evolution of galxies' (SEGAL), and through grant ANR-18-CE31-0006 for project `Galactic Dark Matter' (GaDaMa), the Centre National de la Recherche Scientifique (CNRS) and its SNO \gaia of the Institut des Sciences de l’Univers (INSU), its Programmes Nationaux: Cosmologie et Galaxies (PNCG), Gravitation R\'{e}f\'{e}rences Astronomie M\'{e}trologie (PNGRAM), Plan\'{e}tologie (PNP), Physique et Chimie du Milieu Interstellaire (PCMI), and Physique Stellaire (PNPS), the `Action F\'{e}d\'{e}ratrice \gaia' of the Observatoire de Paris, the R\'{e}gion de Franche-Comt\'{e}, the Institut National Polytechnique (INP) and the Institut National de Physique nucl\'{e}aire et de Physique des Particules (IN2P3) co-funded by CNES;
\item the German Aerospace Agency (Deutsches Zentrum f\"{u}r Luft- und Raumfahrt e.V., DLR) through grants 50QG0501, 50QG0601, 50QG0602, 50QG0701, 50QG0901, 50QG1001, 50QG1101, 50\-QG1401, 50QG1402, 50QG1403, 50QG1404, 50QG1904, 50QG2101, 50QG2102, and 50QG2202, and the Centre for Information Services and High Performance Computing (ZIH) at the Technische Universit\"{a}t Dresden for generous allocations of computer time;
\item the Hungarian Academy of Sciences through the Lend\"{u}let Programme grants LP2014-17 and LP2018-7 and the Hungarian National Research, Development, and Innovation Office (NKFIH) through grant KKP-137523 (`SeismoLab');
\item the Science Foundation Ireland (SFI) through a Royal Society - SFI University Research Fellowship (M.~Fraser);
\item the Israel Ministry of Science and Technology through grant 3-18143 and the Tel Aviv University Center for Artificial Intelligence and Data Science (TAD) through a grant;
\item the Agenzia Spaziale Italiana (ASI) through contracts I/037/08/0, I/058/10/0, 2014-025-R.0, 2014-025-R.1.2015, and 2018-24-HH.0 to the Italian Istituto Nazionale di Astrofisica (INAF), contract 2014-049-R.0/1/2 to INAF for the Space Science Data Centre (SSDC, formerly known as the ASI Science Data Center, ASDC), contracts I/008/10/0, 2013/030/I.0, 2013-030-I.0.1-2015, and 2016-17-I.0 to the Aerospace Logistics Technology Engineering Company (ALTEC S.p.A.), INAF, and the Italian Ministry of Education, University, and Research (Ministero dell'Istruzione, dell'Universit\`{a} e della Ricerca) through the Premiale project `MIning The Cosmos Big Data and Innovative Italian Technology for Frontier Astrophysics and Cosmology' (MITiC);
\item the Netherlands Organisation for Scientific Research (NWO) through grant NWO-M-614.061.414, through a VICI grant (A.~Helmi), and through a Spinoza prize (A.~Helmi), and the Netherlands Research School for Astronomy (NOVA);
\item the Polish National Science Centre through HARMONIA grant 2018/30/M/ST9/00311 and DAINA grant 2017/27/L/ST9/03221 and the Ministry of Science and Higher Education (MNiSW) through grant DIR/WK/2018/12;
\item the Portuguese Funda\c{c}\~{a}o para a Ci\^{e}ncia e a Tecnologia (FCT) through national funds, grants SFRH/\-BD/128840/2017 and PTDC/FIS-AST/30389/2017, and work contract DL 57/2016/CP1364/CT0006, the Fundo Europeu de Desenvolvimento Regional (FEDER) through grant POCI-01-0145-FEDER-030389 and its Programa Operacional Competitividade e Internacionaliza\c{c}\~{a}o (COMPETE2020) through grants UIDB/04434/2020 and UIDP/04434/2020, and the Strategic Programme UIDB/\-00099/2020 for the Centro de Astrof\'{\i}sica e Gravita\c{c}\~{a}o (CENTRA);  
\item the Slovenian Research Agency through grant P1-0188;
\item the Spanish Ministry of Economy (MINECO/FEDER, UE), the Spanish Ministry of Science and Innovation (MICIN), the Spanish Ministry of Education, Culture, and Sports, and the Spanish Government through grants BES-2016-078499, BES-2017-083126, BES-C-2017-0085, ESP2016-80079-C2-1-R, ESP2016-80079-C2-2-R, FPU16/03827, PDC2021-121059-C22, RTI2018-095076-B-C22, and TIN2015-65316-P (`Computaci\'{o}n de Altas Prestaciones VII'), the Juan de la Cierva Incorporaci\'{o}n Programme (FJCI-2015-2671 and IJC2019-04862-I for F.~Anders), the Severo Ochoa Centre of Excellence Programme (SEV2015-0493), and MICIN/AEI/10.13039/501100011033 (and the European Union through European Regional Development Fund `A way of making Europe') through grant RTI2018-095076-B-C21, the Institute of Cosmos Sciences University of Barcelona (ICCUB, Unidad de Excelencia `Mar\'{\i}a de Maeztu’) through grant CEX2019-000918-M, the University of Barcelona's official doctoral programme for the development of an R+D+i project through an Ajuts de Personal Investigador en Formaci\'{o} (APIF) grant, the Spanish Virtual Observatory through project AyA2017-84089, the Galician Regional Government, Xunta de Galicia, through grants ED431B-2021/36, ED481A-2019/155, and ED481A-2021/296, the Centro de Investigaci\'{o}n en Tecnolog\'{\i}as de la Informaci\'{o}n y las Comunicaciones (CITIC), funded by the Xunta de Galicia and the European Union (European Regional Development Fund -- Galicia 2014-2020 Programme), through grant ED431G-2019/01, the Red Espa\~{n}ola de Supercomputaci\'{o}n (RES) computer resources at MareNostrum, the Barcelona Supercomputing Centre - Centro Nacional de Supercomputaci\'{o}n (BSC-CNS) through activities AECT-2017-2-0002, AECT-2017-3-0006, AECT-2018-1-0017, AECT-2018-2-0013, AECT-2018-3-0011, AECT-2019-1-0010, AECT-2019-2-0014, AECT-2019-3-0003, AECT-2020-1-0004, and DATA-2020-1-0010, the Departament d'Innovaci\'{o}, Universitats i Empresa de la Generalitat de Catalunya through grant 2014-SGR-1051 for project `Models de Programaci\'{o} i Entorns d'Execuci\'{o} Parallels' (MPEXPAR), and Ramon y Cajal Fellowship RYC2018-025968-I funded by MICIN/AEI/10.13039/501100011033 and the European Science Foundation (`Investing in your future');
\item the Swedish National Space Agency (SNSA/Rymdstyrelsen);
\item the Swiss State Secretariat for Education, Research, and Innovation through the Swiss Activit\'{e}s Nationales Compl\'{e}mentaires and the Swiss National Science Foundation through an Eccellenza Professorial Fellowship (award PCEFP2\_194638 for R.~Anderson);
\item the United Kingdom Particle Physics and Astronomy Research Council (PPARC), the United Kingdom Science and Technology Facilities Council (STFC), and the United Kingdom Space Agency (UKSA) through the following grants to the University of Bristol, the University of Cambridge, the University of Edinburgh, the University of Leicester, the Mullard Space Sciences Laboratory of University College London, and the United Kingdom Rutherford Appleton Laboratory (RAL): PP/D006511/1, PP/D006546/1, PP/D006570/1, ST/I000852/1, ST/J005045/1, ST/K00056X/1, ST/\-K000209/1, ST/K000756/1, ST/L006561/1, ST/N000595/1, ST/N000641/1, ST/N000978/1, ST/\-N001117/1, ST/S000089/1, ST/S000976/1, ST/S000984/1, ST/S001123/1, ST/S001948/1, ST/\-S001980/1, ST/S002103/1, ST/V000969/1, ST/W002469/1, ST/W002493/1, ST/W002671/1, ST/W002809/1, and EP/V520342/1.
\end{itemize}

The GBOT programme  uses observations collected at (i) the European Organisation for Astronomical Research in the Southern Hemisphere (ESO) with the VLT Survey Telescope (VST), under ESO programmes
092.B-0165,
093.B-0236,
094.B-0181,
095.B-0046,
096.B-0162,
097.B-0304,
098.B-0030,
099.B-0034,
0100.B-0131,
0101.B-0156,
0102.B-0174, and
0103.B-0165;
%
%
and (ii) the Liverpool Telescope, which is operated on the island of La Palma by Liverpool John Moores University in the Spanish Observatorio del Roque de los Muchachos of the Instituto de Astrof\'{\i}sica de Canarias with financial support from the United Kingdom Science and Technology Facilities Council, and (iii) telescopes of the Las Cumbres Observatory Global Telescope Network.

This work made use of software from Postgres-XL (\url{https://www.postgres-xl.org}), Java (\url{https://www.oracle.com/java/}), and TOPCAT/STILTS \citep{2005ASPC..347...29T}.
This research has made use of NASA’s Astrophysics Data System.
This research has made use of the NASA/IPAC Extragalactic Database, which is funded by the National Aeronautics and Space Administration and operated by the California Institute of Technology.
Funding for the Sloan Digital Sky Survey IV has been
provided by the Alfred P. Sloan Foundation, the U.S.
Department of Energy Office of Science, and the Participating
Institutions. SDSS-IV acknowledges support and resources
from the Center for High-Performance Computing at the
University of Utah. The SDSS web site is www.sdss.org.
SDSS-IV is managed by the Astrophysical Research Consortium for the Participating Institutions of the SDSS Collaboration, including the Brazilian Participation Group, the Carnegie
Institution for Science, Carnegie Mellon University, the Chilean
Participation Group, the French Participation Group, HarvardSmithsonian Center for Astrophysics, Instituto de Astrofísica de
Canarias, The Johns Hopkins University, Kavli Institute for the
Physics and Mathematics of the Universe (IPMU)/University of
Tokyo, Lawrence Berkeley National Laboratory, Leibniz Institut
für Astrophysik Potsdam (AIP), Max-Planck-Institut für Astronomie (MPIA Heidelberg), Max-Planck-Institut für Astrophysik
(MPA Garching), Max-Planck-Institut für Extraterrestrische
Physik (MPE), National Astronomical Observatories of China,
New Mexico State University, New York University, University
of Notre Dame, Observatário Nacional/MCTI, The Ohio State
University, Pennsylvania State University, Shanghai Astronomical Observatory, United Kingdom Participation Group, Universidad Nacional Autónoma de México, University of Arizona,
University of Colorado Boulder, University of Oxford, University of Portsmouth, University of Utah, University of
Virginia, University of Washington, University of Wisconsin,
Vanderbilt University, and Yale University.

\end{appendix}
\end{document}